\documentclass[twocolumn,showpacs,preprintnumbers,amsmath,amssymb,superscriptaddress,nofootinbib,english]{revtex4-1}
\usepackage{times,amsmath,amsfonts,amssymb,epstopdf}
\usepackage{graphicx}
\usepackage{dcolumn}
\usepackage{bm}
\usepackage{epsfig}
\usepackage{graphicx}
\usepackage{hyperref}
\usepackage[usenames]{color}
\usepackage{url}
\usepackage[normalem]{ulem}
\usepackage[T1]{fontenc}
\usepackage[dvipsnames]{xcolor}

\newcommand{\tcr}{\textcolor{black}}

\newcommand{\BLED}[1]{\textcolor{black}{#1}}

\newcommand{\mpl}{M_{\rm Pl}}

\newcommand{\rd}{{\rm d}}

\def\ie{{\frenchspacing\it i.e.}}

\def\be{\begin{equation}}
\def\ee{\end{equation}}
\def\ba{\begin{eqnarray}}
\def\ea{\end{eqnarray}}
\begin{document}

\title{Systematic simulations of modified gravity:  chameleon models}

\author{Philippe~Brax}
\email[Email address: ]{philippe.brax@cea.fr}
\affiliation{Institut de Physique Theorique, CEA, IPhT, CNRS, URA 2306, F-91191Gif/Yvette Cedex, France}

\author{Anne-Christine~Davis}
\email[Email address: ]{a.c.davis@damtp.cam.ac.uk}
\affiliation{DAMTP, Centre for Mathematical Sciences, University of Cambridge, Wilberforce Road, Cambridge CB3 0WA, UK}

\author{Baojiu~Li}
\email[Email address: ]{baojiu.li@durham.ac.uk}
\affiliation{Institute for Computational Cosmology, Department of Physics, Durham University, Durham DH1 3LE, UK}

\author{Hans~A.~Winther}
\email[Email address: ]{h.a.winther@astro.uio.no}
\affiliation{Institute of Theoretical Astrophysics, University of Oslo, 0315 Oslo, Norway}

\author{Gong-Bo~Zhao}
\email[Email address: ]{gong-bo.zhao@port.ac.uk}
\affiliation{Institute of Cosmology \& Gravitation, University of Portsmouth, Portsmouth PO1 3FX, UK}
\affiliation{National Astronomy Observatories, Chinese Academy of Science, Beijing, 100012, P.~R.~China}

\date{\today}

\begin{abstract}
In this work we systematically study the linear and nonlinear structure formation in  chameleon theories of modified gravity, using a generic parameterisation which describes a large class of models using only 4 parameters. For this we have modified the $N$-body simulation code {\sc ecosmog} to perform a total of $65$ simulations for different models and parameter values, including the default $\Lambda$CDM. These simulations enable us to explore a significant portion of the parameter space. We have studied the effects of modified gravity on the matter power spectrum and mass function, and found a rich and interesting phenomenology where the difference with the $\Lambda$CDM paradigm cannot be reproduced by a linear analysis even on scales as large as $k\sim0.05$ $h\rm{Mpc}^{-1}$\tcr{, since the latter incorrectly assumes that the modification of gravity depends only on the background matter density}. Our results show that the chameleon screening mechanism is significantly more efficient than other mechanisms such as the dilaton and symmetron, especially in high-density regions and at early times, and can serve as a guidance to determine the parts of  the chameleon parameter space which are cosmologically interesting and thus merit further studies in the future.
\end{abstract}


\maketitle

\section{Introduction}

\label{sect:introduction}

Two plausible alternative explanations to the observed accelerating expansion of our Universe are dynamical dark energy \citep{cst2006} and modified gravity \citep{cfps2011}. In both classes of theories, a scalar field has been used as the most common dynamical origin of the acceleration of  the Universe. This, however, comes at a price: in many theories, especially modified gravity and coupled dark energy theories, dark energy evolves on cosmological time scales only when the scalar field leads to a long range interaction which could violate various gravitational bounds. To avoid this problem, screening mechanisms have been designed to dynamically screen the scalar-mediated fifth force in dense or high-curvature environments.

Screening the effects of a scalar interaction in the presence of matter can be realised in the following ways. Let $\phi_0$ be the environment dependent background configuration and let us expand  the scalar Lagrangian to second order,
\be 
\delta {\cal L}= -\frac{Z(\phi_0)}{2} (\partial \delta \phi)^2 -\frac{m^2(\phi_0)}{2} (\delta\phi)^2 -\beta(\phi_0)\frac{\delta\phi}{M_{\rm Pl}} \delta\rho_m,\nonumber 
\ee
where $Z(\phi_0)$ is the normalisation of the scalar, $m(\phi_0)$ is the mass depending on the background value of the scalar field, and $\beta (\phi_0)$ the coupling to the overdensity $\delta\rho_m$. The fifth force is screened if either $\beta(\phi_0)$ becomes small, or  $m(\phi_0)$ or $Z(\phi_0)$ become large. Hence there are three known screening mechanisms to evade the local gravity constraints: the Vainshtein mechanism \cite{vainshtein} in the Dvali-Gabadadze-Porrati (DGP) \cite{dgp2000} and Galileon \citep{nrt2009,dev2009} models where $Z(\phi_0)$ is large enough to reduce the effective coupling $\beta(\phi_0)/Z^{1/2}(\phi_0)$ below observational levels, the chameleon mechanism \citep{kw2004,ms2007} where the mass $m(\phi_0)$ is large enough to render the range of the scalar interaction smaller than distances probed experimentally and finally in dilaton \citep{bbds2010} and symmetron \citep{hk2010,pospelov} theories the coupling $\beta(\phi_0)$ itself is smaller than observed. These mechanisms all utilise the nonlinearities of the effective scalar Lagrangian to prevent the fifth-force from propagating freely. The nonlinearities for the Vainshtein case stem from the derivative self-couplings of the scalar degree of freedom, while the chameleon, and the dilaton and symmetron use the non-linearities of the potential and the coupling to matter respectively.
In a companion paper \cite{bdlwz2012}, we have presented a systematic study of generic dilaton and symmetron theories in the nonlinear regime of structure formation, using $N$-body simulations based on a unified parameterisation scheme \citep{bdl2011,bdlw2012}. This paper will concentrate on chameleons. In particular, we have generalised the original chameleon models and the parameter space of these new models is analysed making use of $N$-body simulations.

The structure formation in general chameleon models is different from that in GR within the Compton wavelength of the scalar degree of freedom, which is the inverse of the effective mass of the scalar field $m$ \citep{bbdkw2004,bdlw2012,Brax:2005ew}. Indeed the density contrast of matter perturbations increases anomalously there. This implies that  the power spectrum differs from its GR counterpart. It turns out that the scale characterising modified gravity, i.e. the scalar mass now, is large enough to prevent significant effects on linear scales. The main consequences of modified gravity appear in the non-linear regime where numerical methods have to be used. The analysis of the   structure growth of the screened modified gravity models with no higher derivative terms in the Lagrangian, e.g. chameleons,  is rendered easier by the fact that these models can be fully parameterised  by two time-dependent functions, $m(a)$ and $\beta(a)$, i.e. the mass of the scalar field and its coupling to matter as a function of the scale factor. This method works  even on fully nonlinear scales \citep{bdl2011,bdlw2012} \footnote{For other schemes to parameterise modified gravity see \citep{ccm2007,aks2008, jz2008,s2009,fs2010,bfsz2011}. Note however that those schemes are mostly limited to the linear perturbation regime of modified gravity, while the parameterisation here is designed to account for nonlinearities.} where the screening effects on smaller scales, for instance for galaxy halos, can therefore be powerfully analysed.

In this work, we study the nonlinear structure formation using $N$-body simulations for the general chameleon theories that we define here and which are parametrised by two $m(a)$ and $\beta(a)$ functions \cite{bdl2011}. Technically, we use a variant of the {\sc ecosmog} code \citep{ecosmog}, which is based on a public adaptive mesh refinement (AMR) code {\sc ramses} \citep{ramses}, to solve and evolve the $N$-body system.

Although chameleon theories, even in the nonlinear regime, have been studied extensively in the literature (e.g., \cite{lb2007,hs2007,bbds2008,o2008,olh2008,sloh2009,lz2009,zmlhf2010,lz2010,zlk2011,zlk2011b,lh2011,lzk2012,lzlk2012,jblzk2012,lhkzjb2012}), those studies are mostly for specific models in a very restricted parameter space. As an example, simulations for $f(R)$ gravity have thus far only been done for the Hu-Sawicki model \cite{hs2007} which is equivalent to a chameleon theory with the coupling strength $\beta(a)$ fixed to $1/\sqrt{6}$. Our study here, for the first time, allows $\beta(a)$ to have a time evolution\footnote{As we will see below, there are further subtle differences between the time evolutions of $m(a)$ in our chameleon models and the model of \cite{hs2007}.}.Our parameterisation allows us to follow a more systematic approach to vary the different chameleon parameters and study the effects quantitatively. In particular, we find that the chameleon mechanism is considerably more efficient than the dilaton and symmetron mechanisms in restoring GR in high-density regions and at earlier times. Our results here show that linear perturbation theory fails almost whenever it predicts a deviation from $\Lambda$CDM, and point out the portion of the chameleon parameter space that is relevant to cosmology today and in the near future.

The layout of this paper is as follows: in \S~\ref{sect:mod_grav} we review scalar-type theories and show how they can be parameterised simply; in \S~\ref{sect:chameleon} we briefly describe the generalised chameleon model and the possible effects of varying each model parameter; the equations that will be used in the $N$-body simulations are summarised in \S~\ref{sect:nbody_eqns}, various tests of our code are presented in \S~\ref{sec:code_test} and then the cosmological simulations used in this work are discussed in \S~\ref{sect:simulations}; finally we summarise and conclude in \S~\ref{sect:summary}.

To make things clearer, throughout the paper we use the {units $\hbar = c=1$} except  where we use $c$ explicitly. An overbar (subscript $_0$) denotes the background (present-day) value of a quantity and subscript $_\varphi$ means ${\rm d}/{\rm d}\varphi$. $\kappa=8\pi G_N=M_{\rm Pl}^{-2}$, where $M_{\rm Pl}$ is the reduced Planck mass and $G_N$ is Newton's constant, are used interchangeably for convenience.

\section{Scalar-tensor theories of modified gravity}

In this section we briefly describe the essential features of modified gravity theories with a scalar degree of freedom (dof) and how the effects of such a dof can be screened locally to restore general relativity (GR). More detailed descriptions can be found in our previous publications and here we keep the discussion short to make the paper self contained, and familiar readers can skip this section. 

\label{sect:mod_grav}

\subsection{Scalar-tensor theories with screening}

The Einstein-Hilbert action for the scalar field $\varphi$ in a generic scalar-tensor theory has the following form in the Einstein frame,
\begin{eqnarray}\label{eq:action}
S &=& \int {\rm d}^4x\sqrt{-g}\left[\frac{M_{\rm Pl}^2}{2}{R}-\frac{1}{2}\nabla^\mu\varphi\nabla_\nu\varphi- V(\varphi)\right]\nonumber\\
&& + \int {\rm d}^4x \sqrt{-\tilde g} {\cal L}_m(\psi_m^{(i)},\tilde g_{\mu\nu}),
\end{eqnarray}
in which $g$ is the determinant of the metric tensor $g_{\mu\nu}$ and ${ R}$ is the Ricci scalar. We label the $i$th matter field by $\psi_m^{(i)}$. The quantities $\tilde g_{\mu\nu}$ and $\tilde g$ denote respectively the metric tensor in the Jordan frame and its determinant, and they are connected to $g_{\mu\nu}$ and $g$ via the following conformal transformation,
%. More precisely, the excitations of each matter field $%\psi_m^{(i)}$ couple to a metric $\tilde g_{\mu\nu}$ which is related to the Einstein-frame metric $g_{\mu\nu}$ by the conformal rescaling
\be
\tilde g_{\mu\nu}=A^2(\varphi)g_{\mu\nu},~~~~\tilde g=A^8(\varphi) g.
\ee

In the Einstein frame, the equation of motion (EOM) of the scalar field has an extra term because here $\varphi$ explicitly couples to matter, and we get
\be\label{eq:eom}
\Box \varphi= -\beta T + \frac{{\rm d}V}{{\rm d}\varphi},
\ee
in which $T\equiv-\rho+3P$ is the trace of the energy momentum tensor $T^{\mu\nu}$, $\rho, P$ are the energy density and pressure of matter, $\Box\equiv\nabla^\mu\nabla_\mu$ and the coupling strength between $\varphi$ and matter is given by $\beta(\varphi) \equiv \mpl{{\rm d}\ln A}/{{\rm d} \varphi}$.

Eq.~(\ref{eq:eom}) is equivalent to that of a normal quintessence field, with the bare scalar field potential replaced by a new effective potential
\begin{eqnarray}\label{eq:Veff}
V_{\rm eff}(\varphi) &\equiv& V(\varphi) - \big(A(\varphi)-1\big)T.
\end{eqnarray}
In the simplest cases,  $V_{\rm eff}$ has a global minimum in the cosmological background dominated by dust matter for which $P_m=0$ and $T=-\rho_m$. The value of the scalar field at the minimum depends on the actual value of $\rho_m$, i.e., $\varphi_{\rm min}=\varphi_{\rm min}(\rho_m)$. The mass of the scalar field at $\varphi_{\rm min}$, which is defined by
\begin{eqnarray}
m^2 &\equiv& \frac{\rd^2V_{\rm eff}(\varphi)}{\rd\varphi^2}\Bigg|_{\varphi=\varphi_{\rm min}},
\end{eqnarray}
must be positive because an imaginary $m$ can lead to violently unstable evolution of the perturbation of the scalar field.

When matter is described by dust fluid (with radiation negligible) so that there is no anisotropic stress, the line element in the weak-field limit can be expressed as
\be
\rd s^2=-(1+2\Phi) \rd t^2+ (1-2\Phi)\rd{\mathrm x}^2,
\ee
where $\Phi$ is the gravitational potential. This reduces to the modified geodesic equation for matter particles
\be\label{eq:geodesic}
\frac{\rd^2 x^i}{\rd t^2}= -\nabla^i\Big(\Phi+\ln A(\varphi)\Big).
\ee
We can understand Eq.~(\ref{eq:geodesic}) as the motion of a massive particle in an effective gravitational potential
\be
\Psi_{\rm eff}\equiv\Phi+\ln A(\varphi),
\ee
and this is why the theory is considered as a modified gravity theory. 

As an example, let us consider a point mass $M$ embedded in a homogeneous background density as the source of gravity. The effective gravitational potential could be obtained by solving the scalar EOM \cite{bdlwz2012}, as
\be\label{eq:klsol}
\Psi_{\rm eff}= -\Big[1+2\beta(\varphi)^2e^{-m(\varphi)r}\Big]\frac{G_N M}{r}.
\ee
The second term in the brackets represents a Yukawa-type deviation from Newtonian gravity (the fifth force), and this deviation can be of order unity if $mr\lesssim1$ and $\beta\sim{\cal O}(1)$. However, because both $\beta$ and $m$ are functions of the field itself and thus depend on local matter density,  it is possible that near massive bodies nonlinear effects make $\beta(\varphi)\ll1$ or $m^{-1}\ll r$. In such cases, the modification of gravity is strongly suppressed, which helps to evade local constraints on the fifth force. Because the suppression of modified gravity here depends on the massive body itself, we call this {\it self}-screening.

%In addition to the {\it self}-screening described above, the modification of gravity depends on the {\it environment} of the bodies as well. For example, in a high-density background, the scalar field mass $m$ can be very large, which suppresses the deviation from Newtonian gravity as well, according to Eq.~(\ref{eq:klsol}).

{Self}-screening is not the only mechanism to suppress the fifth force in modified gravity theories. Indeed, this suppression often also very strongly depends on the {\it environment} of the body. In the case of chameleon theories, as shown in \citep{bdlw2012}, the fifth force is effectively screened provided that the Newtonian potential $\Phi_N$ at the edge of a massive body follows the relation
\begin{equation}\label{eq:sc}
\vert \varphi_\infty -\varphi_c\vert \ll 2 \beta_\infty M_{\rm Pl} \Phi_N,
\end{equation}
where $\varphi_{c}$ is the minimum of $V_{\rm eff}$ inside the body and $\varphi_{\infty}, \beta_\infty$ are the minimum of $V_{\rm eff}$ and the coupling strength far away. In general, $|\varphi_c|\ll |\varphi_\infty|$, and $\Phi_N$ in Eq.~(\ref{eq:sc}) determines the {\it self}-screening due to the massive body while $\varphi_\infty$ (via also $\beta_\infty$) characterises the {\it environmental}-screening. Note that although $\beta$ is often chosen to be constant in chameleon theories, this does not necessarily have to be the case.

\subsection{Tomography}

\label{subsect:tomography}

As we shall see shortly, a rough estimate of the local constraints on the fifth force indicates that $m^2 \gg H^2$ around the global minimum of $V_{\rm eff}$. The scalar field dynamically tracks  $\varphi_{\rm min}$, around which it oscillates rapidly \cite{bdlw2012}, and the time average $\langle V_{\rm eff}\left(\varphi_{\rm min}\right)\rangle$ then acts as a very slowly-varying cosmological constant. In this simplified case, we can determine the cosmic evolution of the scalar field in terms of $m(a)$ and $\beta(a)$ in background, as \cite{bdl2011,bdlw2012}
\begin{equation}\label{eq:Vofphi}
\varphi(a)=  \frac{3}{\mpl}\int_{a_{\rm ini}}^a \frac{\beta (a)}{a m^2(a)}\rho_m(a)\rd a +\varphi_c,
\end{equation}
where we have assumed $A(\varphi)\doteq1$, as required by the stringent experimental constraints on the time variation of fermion masses, which is proportional to $A$. $\varphi_c$ is the scalar field value at the initial time $a_{\rm ini}$, when the average matter density in the Universe is of the same order as that in typical test bodies in laboratories today. Similarly, we have
\begin{equation}\label{eq:V}
V(a)=V_0 - \frac{3}{\mpl^2}\int_{a_{\rm ini}}^a \frac{\beta^2(a)}{am^2(a)} \rho_m^2(a) \rd a,
\end{equation}
where $V_0=V(a=a_{\rm ini})$.

Given $V(a)$ and $\varphi(a)$, it is straightforward to derive $V$ as a function of $\varphi$, $V(\varphi)$. Similarly, $\beta(\varphi)$ can be reconstructed easily from $\beta(a)$ and $\varphi(a)$. As a result, the full nonlinear dynamics of the theory can be reconstructed elegantly using the background evolutions of $m$ and $\beta$. This `tomography' \cite{bdl2011} has turned out to be very useful as a  generic parameterisation of modified gravity theories and the systematic simulations to study their cosmological implications \cite{bdlwz2012}.

%, and these determine the bare scalar field potential $V(\varphi)$ and the coupling function $\beta(\varphi)$, {\it as functions of $\varphi$}, when $\beta (a)$ and $m(a)$ are given parametrically.
% Hence, we have found that the {\it full} nonlinear dynamics of the theory can be recovered from the knowledge of the {\it time} evolutions of the mass and the coupling to matter since before BBN. 

We can then express the screening condition, Eq.~(\ref{eq:sc}), as
\be\label{eqq}
\int_{a_{\rm in}}^{a_{\rm out}} \frac{\beta (a)}{a m^2(a)}\rho_m(a)\rd a \ll \beta_{\rm out}\mpl^2\Phi_N,
\ee
in which for simplicity we have considered constant matter densities $\rho_{\rm out,in}$ outside and inside the dense body, and $a_{\rm in,out}$ is defined by $\bar{\rho}_{m}(a_{\rm in,out})\equiv\rho_{\rm in,out}$ and $\beta_{\rm out}\equiv\beta(a=a_{\rm out})$.

One can use the fact that the Milky Way must be screened\footnote{\tcr{It happens that the surface Newtonian potential is roughly the same for the Sun and the Galaxy, both $\sim\mathcal{O}(10^{-6})$. So if the Milky Way is not screened to provide environmental screening for the Sun, then the latter will not be self-screened either.}} to make a rough estimate about the screening condition. The averaged matter density inside the Milky Way is $\sim10^6$ times that of the cosmic mean, which implies that $a_{\rm in}\sim 10^{-2}$; its Newtonian potential at its surface is $\Phi_G \sim 10^{-6}$. On the other hand, approximately the environmental matter density for the Milky Way can be taken as close to the cosmic mean\footnote{Clearly, this is only a simplified assumption, because the Milky Way lives in local high-density regions rather than the cosmological background. But here the purpose is only to roughly estimate the possible constraints coming from the Galaxy.}, which gives us $a_{\rm out}\sim 1$. Using these numbers, Eq.~(\ref{eqq}) implies that $m_0/H_0 \gtrsim 10^3$. A similar bound can be deduced from the timing of binary systems \cite{Brax:2013uh} and the distance indicators for stars in dwarf galaxies \cite{Jain:2012tn}. Hence, for a given modified gravity model to be screened locally, the fifth force can only act on scales of and below a few megaparsecs in a cosmological setting. Because $m$ itself is dimensional, in the rest of the paper we shall use the dimensionless quantity
\be
\xi\equiv\frac{H_0}{m_0},
\ee
to parameterise modified gravity theories. $\xi$ is proportional to the range of the fifth force,
\be
\lambda=2998\xi~h^{-1}{\rm Mpc}.
\ee
Even in GR, length scales of order megaparsec are already in the nonlinear regime and cannot be accurately described with linear perturbation theory. The nonlinearity in the equations for modified gravity only makes this situation even worse, and previous experiences \cite{lhkzjb2012} show that linear perturbation theory can be misleading whenever it predicts a deviation from GR. This has motivated us to analyse the large-scale structure formation in the chameleon theory more reliably, using $N$-body simulations.

\section{Generalised chameleon theories}

\label{sect:chameleon}

\subsection{Chameleon theory and its generalisation}

\begin{figure*}
\includegraphics[scale=0.46]{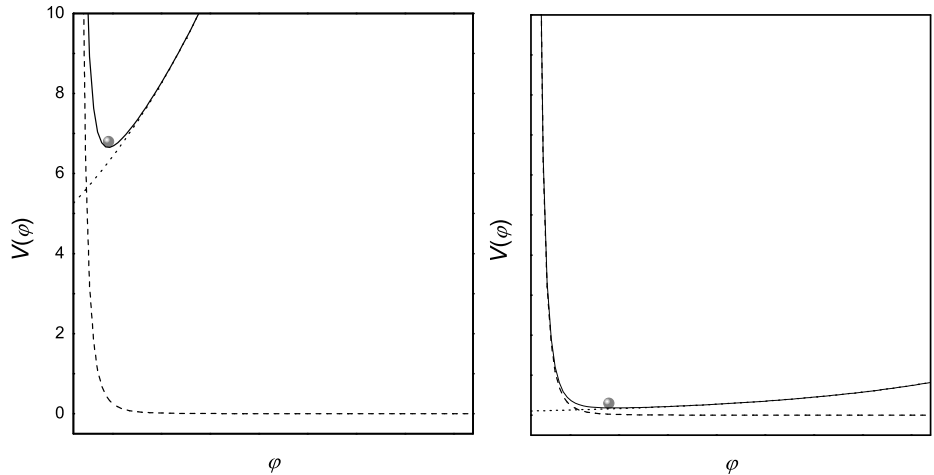}
\caption{Illustration of how the chameleon mechanism works. The dashed, dotted and solid curves are respectively the bare potential $V(\varphi)$ of the chameleon field, the coupling function and the total effective potential $V_{\rm eff}(\varphi)$. {\it Left Panel}: in high matter-density regions the minimum of $V_{\rm eff}(\varphi)$ is very close to $\varphi=0$ and $\vec{\nabla}\varphi$ is small so that the fifth force is strongly suppressed. {\it Right Panel}: in low matter-density regions $\varphi$ and therefore $\vec{\nabla}\varphi$ can be big, and so a nonzero fifth force takes effect in structure formation.} \label{fig:chameleon}
\end{figure*}

In the original chameleon theory proposed in \cite{kw2004, ms2007}, the coupling function and the scalar field bare potential take the following forms respectively:
\begin{eqnarray}
A(\varphi) &=& e^{\beta_0\varphi/M_{\rm Pl}},\\
V(\varphi) &=& V_0\left(\frac{M_{\rm Pl}}{\varphi}\right)^n.
\end{eqnarray}
Here $\beta_0>0$ is a dimensionless model parameter and $V_0$ is a parameter with mass dimension four. The chameleon screening mechanism is graphically illustrated in Fig.~\ref{fig:chameleon}. In high matter-density regions the contribution from the matter coupling to $V_{\rm eff}(\varphi)$ is large and the chameleon field $\varphi$ is trapped in the small-field regime (\ie, $\varphi\rightarrow0$) such that the fifth force, proportional to $\vec{\nabla}\varphi$, is very weak\footnote{\tcr{As an example, in theories with a strong chameleon effect, the scalar field has very small value even in the background and under-dense regions, say ${\varphi}/{\mpl}\in[0,10^{-8}]$. In this case, the variation of $\varphi$ from the inside to the outside of a massive dark matter halo is at most $\sim\mathcal{O}(10^{-8})$, while the variation of the Newtonian potential is typically $\mathcal{O}(10^{-5})$ or even larger, which means the fifth force is much weaker than standard gravity. Indeed, the smallness of $\varphi$ is a generic consequence of the chameleon effect.}}; in low matter-density regions, in contrast, $\varphi$ is big and so is $\vec{\nabla}\varphi$, resulting in a cosmologically interesting fifth force\footnote{One can also understand the suppression of the fifth force in high matter-density regions as a result of the locally very heavy scalar field mass, which characterises the length scale the scalar degree of freedom could propagate without being severely suppressed.}. The essential features of the original chameleon theory include an exponential coupling function $A(\varphi)$ and a runaway potential.

%According to the analysis of \cite{bdl2011}, a coupled scalar field, if heavy enough (namely $m(a)\gg H$), can have its time evolution fully specified by $m(a)$ and $\beta(a)$, both of which are determined as functions of $a$ by the background cosmology.

As discussed above, a coupled scalar field, if heavy enough (namely $m(a)\gg H$), can have its time evolution fully specified by $m(a)$ and $\beta(a)$, both of which are determined as functions of $a$ by the background cosmology. For the chameleon theory listed in \cite{bdl2011}, it has been shown that
\begin{eqnarray}\label{eq:m_original}
m(a) &=& m_0a^{-r},\\
\label{eq:beta_original}\beta(a) &=& \beta_0.
\end{eqnarray}
where $r>0$.

As a straightforward generalisation of the chameleon idea, in this paper we shall consider a power-law form of both $m(a)$ (as in Eq.~(\ref{eq:m_original})) and $\beta(a)$:
\begin{eqnarray}\label{eq:beta_new}
\beta(a) &=& \beta_0a^{-s},
\end{eqnarray}
where $s$ is a new model parameter to describe the generalised chameleon theory. Using the tomographic mapping discussed above, we find that
\begin{eqnarray}\label{eq:varphi_of_a}
\frac{\varphi(a)}{M_{\rm Pl}} &=& \frac{\varphi_i}{M_{\rm Pl}} + \int^a_{a_i}\frac{\beta(a)}{am^2(a)}\kappa\rho_m(a)da\nonumber\\
&=& \frac{\varphi_i}{M_{\rm Pl}} + 9\Omega_m\beta_0\xi^2\frac{1}{2r-s-3}\left[a^{2r-s-3}-a^{2r-s-3}_i\right],\nonumber
\end{eqnarray}
where we have used a subscript $_i$ to denote the value of a quantity at the initial time $a_i$, and $\xi=H_0/m_0$ as defined above. As we take the limit $a_i\rightarrow0$, the chameleon field is driven to $\varphi\rightarrow0$ and the above equation reduces to
\begin{eqnarray}\label{eq:varphi_of_a}
\frac{\varphi(a)}{M_{\rm Pl}} &=& \frac{9}{2r-s-3}\Omega_m\beta_0\xi^2a^{2r-s-3}.
\end{eqnarray}

In order to study the nonlinear evolution of $\varphi$, we have to know $V_\varphi(\varphi)$, where a subscript $_\varphi$ denotes derivative with respect to $\varphi$, which governs the dynamics of the scalar field (see the $N$-body equations below).  For this we find
\begin{eqnarray}
\kappa V_\varphi &=& \frac{d(\kappa V(a))}{d a}\frac{da}{d\varphi}\nonumber\\
&=& \label{eq:V_varphi_of_a}-3\Omega_m\beta_0H_0^2a^{-s-3}\\
&=& \label{eq:V_varphi_of_varphi}-3\Omega_m\beta_0H_0^2\left[\frac{9\Omega_m\beta_0}{2r-s-3}\right]^{\frac{3+s}{2r-s-3}}\left[\frac{\xi^2M_{\rm Pl}}{\varphi}\right]^{\frac{3+s}{2r-s-3}},\
\end{eqnarray}
in which Eq.~(\ref{eq:V_varphi_of_a}) can be used in background cosmology and linear perturbation analysis to replace $V_\varphi$. As $\beta_0,\Omega_m$ and $\xi^2$ are all positive, to make sure that the quantities in Eq.~(\ref{eq:V_varphi_of_varphi}) are well defined we will require $2r-s-3>0$ and $\varphi>0$ in what follows\footnote{Otherwise the terms in the brackets could be negative, making the power-law function ill defined.}. \tcr{We also require that $r\geq2$, since $H^2\propto a^{-3}$ during the matter-dominated era and $H^2\propto a^{-4}$ in the radiation era, so that one may have $H^2>m^2$ at early times if $r<2$.}

With $V_\varphi(\varphi)$, one could easily integrate to obtain $V(\varphi)$ analytically and we have
\begin{eqnarray}
\kappa V(\varphi) &=& \kappa V_0 - \frac{27\Omega_m^2\beta_0^2\xi^2H_0^2}{2r-2s-6}\left[\frac{2r-s-3}{9\Omega_m\beta_0\xi^2}\frac{\varphi}{M_{\rm Pl}}\right]^{\frac{2r-2s-6}{2r-s-3}} \ \
\end{eqnarray}
for $r-s\neq3$ and
\begin{eqnarray}
\kappa V(\varphi) &=& \kappa V_0 - \frac{27}{r}\Omega_m^2\beta_0^2\xi^2H_0^2\log\left[\frac{r}{9\Omega_m\beta_0\xi^2}\frac{\varphi}{M_{\rm Pl}}\right]\ \
\end{eqnarray}
for $r-s=3$. The perturbation of the dark energy density, $V(\varphi)-V(\bar{\varphi})$, appears in the source for the Poisson equation (see below), but it is generally very small and can be safely neglected.

Given $\varphi(a)$ and $\beta(a)$, it is straightforward to find $\beta(\varphi)$. For our parameterisation using $r,s$, we find that
\begin{eqnarray}\label{eqb}
A(\varphi) &\doteq& 1+\left[\beta_0\left(\frac{9\Omega_m\beta_0\xi^2}{2r-s-3}\right)^{\frac{s}{2r-s-3}}\frac{2r-s-3}{2r-2s-3}\right]\nonumber\\
&&\times\left[\frac{\varphi}{M_{\rm Pl}}\right]^{\frac{2r-2s-3}{2r-s-3}}
\end{eqnarray}
and
\begin{eqnarray}\label{eqa}
\beta(\varphi) &\doteq& \beta_0\left[\frac{2r-s-3}{9\Omega_m\beta_0\xi^2}\frac{\varphi}{M_{\rm Pl}}\right]^{-\frac{s}{2r-s-3}}.
\end{eqnarray}
As a result, both $V_{\varphi}$ and $\beta$ are power-law functions of $\varphi$.

\subsection{Effects of varying chameleon parameters}

\label{subsect:effect_param}

As shown above, our generalised chameleon theory is specified by four model parameters, namely, $\beta_0, r, s$ and $\xi$. The effect of varying these four parameters on the structure formation can be understood without solving the system explicitly.

The parameter $\beta_0$, which is  the coupling strength at $z=0$, controls the overall amplitude of the coupling throughout the entire evolution history. The larger $\beta_0$ is, the stronger the fifth force is, thus the strong clustering of matter relative to that in $\Lambda$CDM (in which $\beta_0=0$).

The parameter $r$, which is the power index of $m(a)$, determines the time evolution of the effective mass of the scalar field without changing $m_0$, which is the mass at $z=0$. The smaller $r$ is ($r>0$), the lighter the scalar field is at $z>0$, and so the longer the range of the fifth force is. Due to the tomography mapping, this also implies that the scalar field is less heavy in high-density regions, leading to a weaker chameleon screening and a stronger clustering of matter. \tcr{Recall from the above that we restrict ourselves to $r\geq2$.}
%Notice that we cannot have $r<0$ since otherwise the fifth force is {\it stronger} in {\it high}-density regions and we have an {\it anti}-chameleon effect.

The parameter $s$, which is the power index of $\beta(a)$, determines the time evolution of the coupling function. The more negative $s$ is, the weaker the coupling between matter and the scalar field at $z>0$ becomes; because of the tomography relation, this also means a stronger suppression of the fifth force in high-density regions, and therefore weaker matter clustering. \tcr{Note that here we restrict ourselves to $s<0$ to avoid the {\it anti}-chameleon effect: this can be seen by looking at Eq.~(\ref{eqa}), which shows that the coupling is stronger in high density regions, or Eq.~(\ref{eq:beta_new}), which shows that the coupling is stronger at earlier times. The situation is worse if $r-3/2\leq s\leq 2r-3$, which implies that $A(\varphi)$ decreases with $\varphi$ by Eq.~(\ref{eqb}), and there is no longer any minimum for $V_{\rm eff}$.}

The parameter $\xi$, which is simply $H_0/m_0$, essentially sets the effective mass $m$ of the scalar field (and thus the effective range of the fifth force) at $z=0$. In all the chameleon simulations we study in this work, $\xi\ll1$. The larger $\xi$ is, the lighter the scalar field is and the stronger the fifth force becomes.

In what follows, we will find that the $N$-body simulations confirm this analysis and also quantify these effects.

\section{The $N$-body Equations}

\label{sect:nbody_eqns}

This section serves to introduce the $N$-body Poisson and chameleon equations for the sake of completeness. For this we list the equations to be solved and describe the code units used in our simulations, both of which can be found in \cite{ramses,ecosmog,bdlwz2012}. 

\subsection{Simplified field equations}

The relevant equations which determine the dynamics of the chameleon and gravity fields are
\begin{eqnarray}
\label{eq:newton}\nabla^2\Phi &\approx& 4\pi G\left(\rho_m-\bar{\rho}_m\right),\\
\label{eq:sf}c^2\nabla^2\varphi &\approx& V_\varphi(\varphi)-V_\varphi(\bar{\varphi})+A_\varphi(\varphi)\rho_m-A_\varphi(\bar{\varphi})\bar{\rho}_m,\\
\label{eq:particle}\frac{{\rm d}^2\vec{r}}{\rd t^2} &=& -\vec{\nabla}\Phi - c^2\BLED{\beta(\varphi)}\vec{\nabla}\varphi-\BLED{\beta(\varphi)}\dot{\varphi}\frac{\rd \vec{r}}{\rd t},
\end{eqnarray}
where we work in the quasi-static limit by dropping all terms involving time derivatives.

The validity of the quasi-static approximation was tested explicitly in \cite{o2008}, which compared the time and spatial derivatives and found that the former is indeed negligible. Note that, rigorously speaking, \cite{o2008} only tested that $\frac{d}{dt}\langle\varphi\rangle$ is negligible, where $\langle\varphi\rangle$ is the scalar field value averaged over the quick oscillations, rather than $\dot{\varphi}$ itself, which can be as large as $|\vec{\nabla}\varphi|$ due to the oscillations. The oscillations themselves, however, largely cancel out and it is the averaged effect that we observe -- in this sense we believe that the test of \cite{o2008} is accurate. We have checked, using our linear perturbation code, that the effects on cosmological observables (such as $\sigma_8$) differ by less than $\sim0.1\%$ in the two cases where we respectively follow the oscillations accurately and average over them \cite{bdlw2012}.

It is tempting to try to solve the full time-dependent scalar field EOM \cite{lm2013} in modified gravity simulations, but notice that to follow the time evolution one has to resolve the oscillations very well. It does not seem so difficult at the background level, where $m_0/H_0\sim10^3$, meaning that to accurately resolve the oscillations one needs a factor of $\mathcal{O}(10)\times\mathcal{O}(10^3)\sim\mathcal{O}(10^4)-\mathcal{O}(10^5)$ coarse time steps. However, even in a {\it mildly} high-density region one could have $m_{\rm local}/H_0\sim10^6$, requiring $\mathcal{O}(10^7)-\mathcal{O}(10^8)$ time steps to accurately follow the time evolution during the course of an $N$-body simulation.  For comparison, the simulations in this paper uses a few hundred coarse time-steps so fully solving the EOM represents a huge increase in the computational cost of a simulation. Using fewer time steps would mean that some sort of average has been done implicitly, in the same sense as it is done in the quasi-static approximation.

A full treatment of this issue is beyond the scope of this work, but rigorous test of the quasi-static approximation for modified gravity theories is something we plan to pursue in the future.

\subsection{Code units}

The code units  are based on (but not exactly) the supercomoving coordinates of \cite{ms1998}. They can be summarised as follows (tilded quantities are expressed in the code unit):
\begin{eqnarray}
\tilde{x}\ =\ \frac{x}{aB},\ \ \ d\tilde{t}\ =\ H_0\frac{dt}{a^2},\ \ \ \tilde{c}\ =\ \frac{c}{BH_0},\nonumber\\
\tilde{\Phi}\ =\ \frac{a^2\Phi}{(BH_0)^2},\ \ \ \tilde{\rho}\ =\ \frac{\rho a^3}{\rho_c\Omega_m},\ \ \ \tilde{v}\ =\ \frac{av}{BH_0},\nonumber
\end{eqnarray}
where $x$ is the physical coordinate,  $t$ is the physical time, $c$ is the speed of light, $\rho_c$ the critical density at present, $\Omega_m$ today's fractional energy density for matter, $v$ the particle velocity and $\Phi$ the Newtonian potential. Besides, $B$ is the simulation box size in unit of $h^{-1}$Mpc. The average matter density is $\tilde{\bar{\rho}}=1$ in the code unit. Note that all these code quantities are dimensionless.

\subsection{The discrete equations}

In cosmological simulations, the chameleon field $\varphi$ is generally extremely small (\ie, $\varphi/M_{\rm Pl}\ll1$) and must be positive to make the logarithmic in Eq.~(\ref{eq:V_varphi_of_varphi}) well defined. To prevent the numerical value of $\varphi$ from becoming negative during the computation and therefore causing divergence problems, we follow \cite{o2008, lz2009, lz2010} to define a new variable $u=\log(\varphi/M_{\rm Pl})$ instead of using $\varphi$ itself. Throughout the cosmic evolution and from one spatial position to another, $\varphi$ can change by several orders of magnitude, but $|u|$ remains $\mathcal{O}(1\sim10)$, making the numerical problems easier to avoid when using $u$.

Using the quantities defined above, the Poisson equation, Eq.~(\ref{eq:newton}), can be written as
\begin{eqnarray}\label{eq:newton_nbody}
\tilde{\nabla}^2\tilde{\Phi} &\approx& \frac{3}{2}\Omega_ma\left(\tilde{\rho}-1\right),
\end{eqnarray}
and, after some manipulation, the chameleon equation of motion Eq.~(\ref{eq:sf}) reads
\begin{eqnarray}\label{eq:sf_nbody}
\tilde{\nabla}^2e^u &\approx& \frac{3}{\tilde{c}^2}\Omega_m\beta_0\tilde{\rho}\left[\frac{e^u}{\bar{\varphi}}\right]^{-\frac{s}{2r-s-3}}a^{-1-s}\\
&&-\frac{3}{\tilde{c}^2}\Omega_m\beta_0\left[\frac{e^u}{\bar{\varphi}}\right]^{-\frac{3+s}{2r-s-3}}a^{-1-s}.\nonumber
\end{eqnarray}

%From here on we will only use variables expressed using the code unit, and the tildes will be dropped for simplicity for all quantities except $\tilde{\rho}$.

Before being implemented into the $N$-body code, the above equations must be discretised. For the Poisson equation this is straightforward and we have
\begin{eqnarray}\label{eq:newton_discrete}
\frac{1}{h^2}\big[\tilde{\Phi}_{i+1,j,k}+\tilde{\Phi}_{i-1,j,k}+\tilde{\Phi}_{i,j+1,k}+\tilde{\Phi}_{i,j-1,k}+\tilde{\Phi}_{i,j,k+1}\nonumber\\
+\tilde{\Phi}_{i,j,k-1}-6\tilde{\Phi}_{i,j,k}\big] = \frac{3}{2}\Omega_ma\left(\tilde{\rho}_{i,j,k}-1\right),\ \ \
\end{eqnarray}
where $\tilde{\Phi}_{i,j,k}$ denotes the value of $\tilde{\Phi}$ in the $(i,j,k)$-th cell of the simulation grid. The discretised nonlinear chameleon equation can be obtained in a similar way though it involves longer derivation,
\begin{eqnarray}\label{eq:sf_discrete}
L^h(u_{i,j,k}) &=& 0,
\end{eqnarray}
with the operator $L^h(u_{i,j,k})$ defined as
\begin{widetext}
\begin{eqnarray}\label{eq:op}
L^h(u_{i,j,k}) &\equiv& \frac{1}{h^2}\left[b_{i+\frac{1}{2},j,k}u_{i+1,j,k}-u_{i,j,k}\left(b_{i+\frac{1}{2},j,k}+b_{i-\frac{1}{2},j,k}\right)+b_{i-\frac{1}{2},j,k}u_{i-1,j,k}\right]\nonumber\\
&&+\frac{1}{h^2}\left[b_{i,j+\frac{1}{2},k}u_{i,j+1,k}-u_{i,j,k}\left(b_{i,j+\frac{1}{2},k}+b_{i,j-\frac{1}{2},k}\right)+b_{i,j-\frac{1}{2},k}u_{i,j-1,k}\right]\nonumber\\
&&+\frac{1}{h^2}\left[b_{i,j,k+\frac{1}{2}}u_{i,j,k+1}-u_{i,j,k}\left(b_{i,j,k+\frac{1}{2}}+b_{i,j,k-\frac{1}{2}}\right)+b_{i,j,k-\frac{1}{2}}u_{i,j,k-1}\right]\nonumber\\
&&+\frac{3}{\tilde{c}^2}\Omega_m\beta_0\left[\frac{e^{u_{i,j,k}}}{\bar{\varphi}}\right]^{-\frac{3+s}{2r-s-3}}a^{-1-s}-\frac{3}{\tilde{c}^2}\Omega_m\beta_0\tilde{\rho}\left[\frac{e^{u_{i,j,k}}}{\bar{\varphi}}\right]^{-\frac{s}{2r-s-3}}a^{-1-s}.
\end{eqnarray}
\end{widetext}
Here $b\equiv\partial e^u/\partial u = e^u$,
\begin{eqnarray}
b_{i+\frac{1}{2},j,k} &\equiv& \frac{1}{2}\left(b_{i+1,j,k}+b_{i,j,k}\right),\nonumber\\
b_{i-\frac{1}{2},j,k} &\equiv& \frac{1}{2}\left(b_{i,j,k}+b_{i-1,j,k}\right),~\cdots\nonumber
\end{eqnarray}
and $h$ is the cell size in the simulation grid \footnote{Note that $h$ is also used in this paper to denote $H_0/(100~\mathrm{km/s/Mpc})$, but there should be no confusion since it is easy to understand its actual meaning based on the context.}.

In our simulation, Eq.~(\ref{eq:sf_discrete}) will be solved using the Newton Gauss-Seidel relaxation method, described as
\begin{eqnarray}\label{eq:ngs}
u^{h,\rm new}_{i,j,k} &=& u^{h,\rm old}_{i,j,k}-\frac{L^h\big(u^{h,\rm old}_{i,j,k}\big)}{\frac{\partial L^h\left(u^{h,\rm old}_{i,j,k}\right)}{\partial u^{h,\rm old}_{i,j,k}}},
\end{eqnarray}
where
\begin{widetext}
\begin{eqnarray}\label{eq:dop}
\frac{\partial L^h\left(u_{i,j,k}\right)}{\partial u_{i,j,k}} &=& \frac{\tilde{c}^2}{2h^2}b_{i,j,k}\big[u_{i+1,j,k}+u_{i-1,j,k}+u_{i,j+1,k}+u_{i-1,j,k}+u_{i,j,k+1}+u_{i,j,k-1}-6u_{i,j,k}\big]\nonumber\\
&&-\frac{\tilde{c}^2}{2h^2}\big[b_{i+1,j,k}+b_{i-1,j,k}+b_{i,j+1,k}+b_{i,j-1,k}+b_{i,j,k+1}+b_{i,j,k-1}+6b_{i,j,k}\big]\nonumber\\
&&+3\Omega_mA_2e^u_{i,j,k}a^2\left[a^{2r-3}+\frac{2r-3}{s}\frac{u_{i,j,k}}{\bar{\varphi}}\right]^{-\frac{3}{2r-3}}-\frac{1}{\xi^2}e^{u_{i,j,k}}a^2\left[a^{2r-3}+\frac{2r-3}{s}\frac{u_{i,j,k}}{\bar{\varphi}}\right]^{-\frac{2r}{2r-3}}\nonumber\\
&&-3\Omega_mA_2\tilde{\rho}a^{-2}e^{u_{i,j,k}}.
\end{eqnarray}
\end{widetext}

%In practice, Eqs.~(\ref{eq:op}) and (\ref{eq:dop}) should be modified at the boundaries of refinements for the multigrid implementation, as is the case of the Poisson equation. \cite{ecosmog} gives a detailed review of all the technical details involved in the $N$-body code implementation, and interested readers are referred to that paper.

\section{Code tests}

\label{sec:code_test}

\begin{table}
\label{tab:test_models} \caption{The parameter values for the seven models used in the code test.}
\begin{tabular}{@{}lcccc}
\hline\hline
parameter\ \ \ \  & $\beta_0$\ \ \ \  & $r$\ \ \ \  & $s$\ \ \ \  & $\xi$\ \ \ \ \\
\hline
model a\ \ \ \  & $0.5$\ \ \ \ & $3.0$\ \ \ \  & $0.0$\ \ \ \  & $0.001$\ \ \ \ \\
model b\ \ \ \  & $1.0$\ \ \ \  & $3.0$\ \ \ \  & $0.0$\ \ \ \  & $0.001$\ \ \ \ \\
model c\ \ \ \  & $0.5$\ \ \ \  & $4.0$\ \ \ \  & $0.0$\ \ \ \  & $0.001$\ \ \ \ \\
model d\ \ \ \  & $0.5$\ \ \ \  & $3.0$\ \ \ \  & $-1.0$\ \ \ \  & $0.001$\ \ \ \ \\
model e1\ \ \ \  & $0.5$\ \ \ \  & $3.0$\ \ \ \  & $0.0$\ \ \ \  & $0.0005$\ \ \ \ \\
model e2\ \ \ \  & $0.5$\ \ \ \  & $3.0$\ \ \ \  & $0.0$\ \ \ \  & $0.002$\ \ \ \ \\
model e3\ \ \ \  & $0.5$\ \ \ \  & $3.0$\ \ \ \  & $0.0$\ \ \ \  & $0.005$\ \ \ \ \\
\hline\hline
\end{tabular}
\end{table}

To make sure that our code works properly, we performed a number of code tests, which are described in this section. We tested the code for 7 models by varying the 4 parameters for the generalised chameleon model, namely $\beta_0,r,s$ and $\xi$, and these are summarised in table~\ref{tab:test_models}. Throughout this section we adopt the unit $M_{\rm Pl}=1$.

\subsection{Homogeneous matter density field}

In a homogeneous matter density field, the chameleon field $\varphi$ must take a constant value given by
\begin{eqnarray}\label{eq:varphi_background}
\varphi &=& \frac{9}{2r-s-3}\Omega_m\beta_0\xi^2a^{2r-s-3}.
\end{eqnarray}
Our first test, therefore, is to fix the matter density field on the simulation grid, make a random initial guess about the values of $u$, let the Newton Gauss-Seidel relaxation iterate for a few steps, and see if $u$ approaches $\log\varphi$ ($\varphi$ given in the above equation) in all grid cells. 

%Thus, as a simplest test of the chameleon equation solver, one could show that in such a homogeneous field, given some random initial guess of $u$ on the cells of the simulation mesh, after a reasonable number of Gauss-Seidel relaxation sweeps, the solutions all converge to the above background value. Such simple test have been used previously in \cite{bbdls2011,ecosmog,dlmw2012} to show that the solver for extra degrees of freedom works correctly.

\begin{figure}
\includegraphics[scale=0.36]{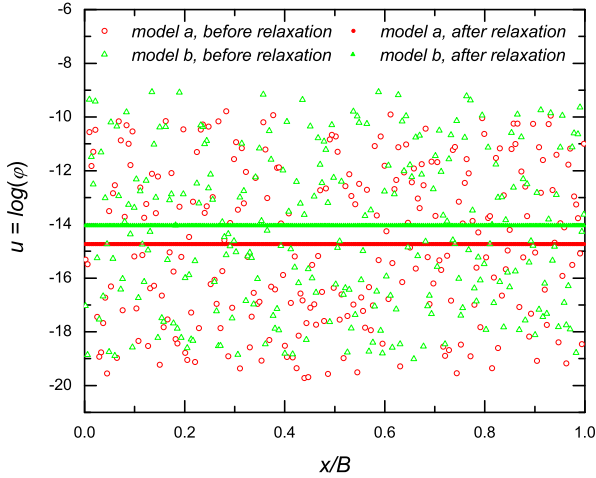}
\caption{(Colour online) Homogeneous matter field test. Shown are the values of the scalar field $\varphi$ in all cells along the $x$-direction, with $y=z=0$. To make the plot clearer, only results for models a and b are shown. Open symbols represent the initial guess and filled symbols of the same shape and colour represent the numerical solutions. Please note that, instead of $\varphi$ itself, shown here is $u=\log(\varphi)$, which is what the code outputs directly. The thick solid lines are the analytical solutions specified in Eq.~(\ref{eq:varphi_background}) for the two models, and they have the same colour as the corresponding numerical solutions.} \label{fig:test_const_dens}
\end{figure}

We did this test for 3 out of the 5 models described in Table~\ref{tab:test_models} at $a=1.0$, as shown in Fig.~\ref{fig:test_const_dens}, in which we plotted the values of $u$ in all cells in the $x$-direction, both before (open symbols) and after (filled symbols) the relaxation. We could see clearly a good agreement between the numerical solutions (filled symbols) and analytic results (the horizontal lines). We also did the test at several values of $a<1.0$ and found similar agreements, but these are now shown here for clarity.

\subsection{Point mass}

Our second test makes use of the simplest spherically symmetric density field, a point mass at the origin \cite{o2008,bbdls2011,ecosmog}, In which case there is an exact analytic solution to $\varphi$ (or equivalently $u$)  under certain simplifications.

%As the second test of our chameleon equation solver, consider the solution of $\varphi$ around a point mass at the origin, in which case we have an analytical solution which is accurate except for the regions very close to the mass..

The said density configuration can be constructed \cite{o2008} as
\begin{equation}
\label{eq:point_mass}
\delta_{i,j,k} = \left\{%
\begin{array}{ll}
10^{-4}\left(N^3-1\right), & \hbox{$i=j=k=0$;} \\
-10^{-4}, & \hbox{otherwise.} \\
\end{array}%
\right.
\end{equation}
in which we have defined $\delta_{i,j,k}\equiv\tilde{\rho}_{i,j,k}-1$. The analytical solution can be obtained by solving the equation
\begin{eqnarray}\label{eq:linearised_eqn}
\nabla^2\delta\varphi &=& m^2\delta\varphi
\end{eqnarray}
away from the origin (where the point mass is), in which the mass of the chameleon dof, $\delta\varphi\equiv\varphi-\bar{\varphi}$, is given by $m^2=\xi^2H_0^2$, and by doing that we found
\begin{eqnarray}\label{xxx}
\delta\varphi &\propto& \frac{1}{r}\exp(-mr),
\end{eqnarray}
in which $r$ is the distance to the origin.

\begin{figure}
\includegraphics[scale=0.36]{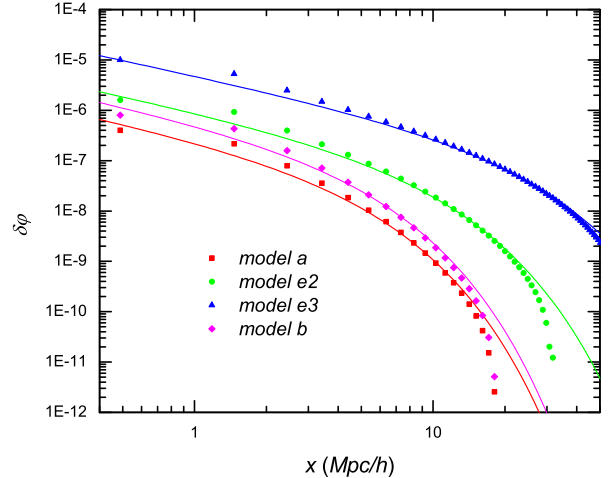}
\caption{(Colour online) Point mass tests. Shown here as filled symbols are the values of $\delta\varphi=\varphi-\bar{\varphi}$ away from a point particle constructed as described in Eq.~(\ref{eq:point_mass}), along the $x$-direction. To make the plot less busy, we only show the results for 4 out of the 7 test models listed in Table~\ref{tab:test_models} (more details in the legends). We use solid curves for the analytical solutions of Eq.~(\ref{xxx}), which are good approximations if the distance to the point mass is not too small.} \label{fig:test_point_mass}
\end{figure}

In this test, the simulation box we chose has a length of $250h^{-1}$Mpc and 256 cells on each side, and we tested all the 5 models at $a=1.0$. Fig.~\ref{fig:test_point_mass} compares the numerical solutions (symbols) of $\delta\varphi$ in the $x$-direction to the analytical predictions (solid curves) given in Eq.~(\ref{xxx}), and it is clear that they agree very well for all tested models. We stress that the discrepancies on large and small values of $r$ are not indications of the code's failure -- the former is because the magnitude of $\delta\varphi$ is already at the level of discretisation error (which limits the code's ability to make the solution more accurate by further relaxation iterations, and this can be seen from the fact that the discrepancy happens at the same value of $\delta\varphi$ for all models), and the latter is because of the fact that in deriving Eq.~(\ref{eq:linearised_eqn}) one artificially linearises a nonlinear equation \cite{o2008}.

%The discrepancies at small $x$ is because the linearisation procedure in deriving  is not accurate and the discrepancy at big $x$ is because the size of $\delta\varphi$ has reached to the level of discretisation error and one cannot get more accurate \cite{o2008}.

\subsection{Sine density field}

The next two tests make use of one-dimensional density configurations, which are obtained by eliminating the $y$- and $z$-dependences of the density field. Starting from a given 1D solution to $\varphi$, we substituted it into the chameleon EOM to find the desired density field, and then used this density field in the numerical code to solve for $\varphi$ and compare with that original input.

The first such test uses a sine density field as first introduced in \cite{o2008}, which in our code units can be written as
\begin{eqnarray}\label{eq:sine_dens}
\tilde{\rho}(x) &=& \frac{\tilde{c}^2a^{1+s}\bar{\varphi}}{\Omega_m\beta_0}\frac{(2\pi)^2}{3}\sin(2\pi x)\left[2-\sin(2\pi x)\right]^{\frac{s}{2r-s-3}}\nonumber\\
&&+\left[2-\sin(2\pi x)\right]^{-\frac{3}{2r-s-3}},
\end{eqnarray}
in which $x$ is rescaled by $B$ and so $x\in[0,1]$. The corresponding $\varphi$ field which is associated wth this density field is
\begin{eqnarray}\label{yyy}
\varphi(x) &=& \bar{\varphi}\left[2-\sin(2\pi x)\right].
\end{eqnarray}

\begin{figure}
\includegraphics[scale=0.36]{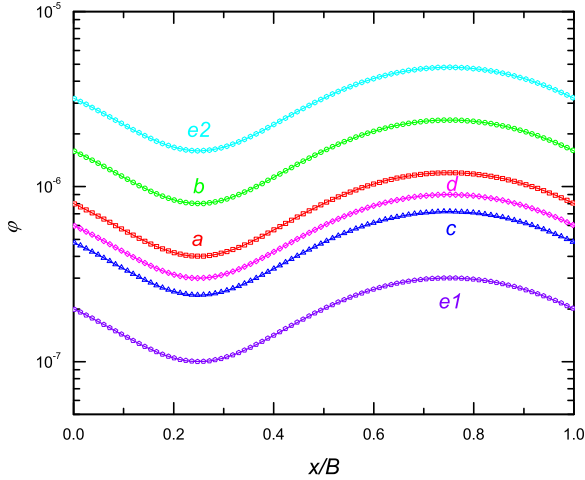}
\caption{(Colour online) Tests with sine density fields, for 6 out of the 7 test models (a, b, c, d, e1, e2) at $a=1.0$ (see the legends). Shown here are the numerical solutions (filled symbols) of $\varphi$ along the $x$-direction in the sine-type density field described in Eq.~(\ref{eq:sine_dens}), and the analytical results of Eq.~(\ref{yyy}) (solid curves of the same colour as the symbols). For this test the simulation box size is $250h^{-1}$Mpc and $x$ is rescaled to make $x/B\in[0,1]$. The simulation mesh has $256^3$ cells.} \label{fig:test_sine_dens}
\end{figure}

We did this test for 6 models listed in Table~\ref{tab:test_models} at $a=1.0$ and the results are shown in Fig.~\ref{fig:test_sine_dens}. As expected, there is a good agreement between the numerical solutions (open symbols) and the analytical results (filled symbols).

%Fig.~\ref{fig:test_sine_dens} shows the test results for the sine density field given above, at $a=1.0$ for six models listed in Table~\ref{tab:test_models}. It is seen that the numerical solutions (empty symbols) agree with the analytical predictions (solid curves) very well.

\subsection{Gaussian density field}

The second test that makes us of a 1D density field assumes a Gaussian-type solution to $\varphi$, given by
\begin{eqnarray}\label{eq:varphi_gaussian_dens}
\varphi &=& \bar{\varphi}\left[1-\alpha\exp\left(-\frac{(x-0.5)^2}{W^2}\right)\right],
\end{eqnarray}
where $W$, $\alpha$ are numerical constants which are used to specify the width and height of the Gaussian function. As before, $x$ is scaled by the boxsize so that $x\in[0,1]$. Note that the Gaussian function in $\varphi(x) $peaks at $x=0.5$ while at $x\rightarrow0$ or $x\rightarrow1$ we have $\varphi\rightarrow\bar{\varphi}$. Also, $\alpha\rightarrow1$ makes $|\varphi|$ very small at $x=0.5$.

The density field which is associated to the above solution to $\varphi(x)$ is 
\begin{eqnarray}\label{eq:gaussian_dens}
\tilde{\rho}(x) &=& \frac{\tilde{c}^2a^{1+s}}{3\Omega_m\beta_0}\frac{2\alpha}{W^2}\frac{\exp\left[-\frac{(x-0.5)^2}{W^2}\right]\left[1-2\frac{(x-0.5)^2}{W^2}\right]}{1-\alpha\exp\left[-\frac{(x-0.5)^2}{W^2}\right]^{\frac{2r-s-3}{s}}}\nonumber\\
&&+\left[1-\alpha e^{-\frac{(x-0.5)^2}{W^2}}\right]^{-\frac{3}{2r-3}}a^3.
\end{eqnarray}
Notice that such a density field is not exactly periodic at the edges of the simulation box, but given that $W$ is small enough, $\tilde{\rho}\rightarrow0$ at the box edges and periodic boundary conditions are approximately satisfied.

%where again $x$ has been scaled to code units so that $x\in[0,1]$, $W$, $\alpha$ are numerical constants which respectively specify the width and height of the density field, which obviously peaks at $x=0.5$.

\begin{figure}
\includegraphics[scale=0.36]{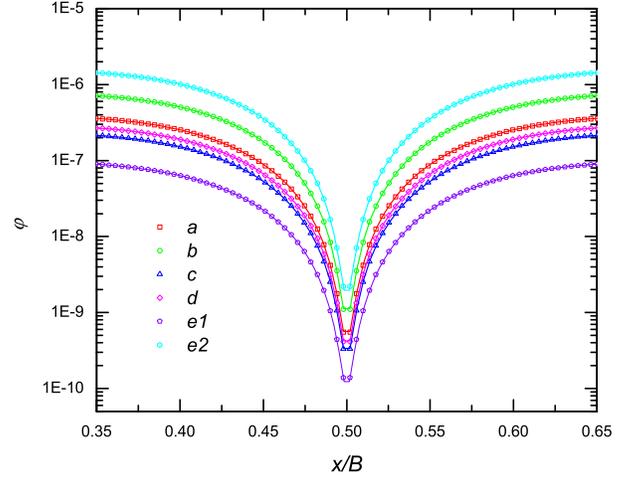}
\caption{(Colour online) Tests with Gaussian density fields, for 6 out of the 7 test models (a, b, c, d, e1, e2) at $a=1.0$ (see the legends for details). Shown here are the numerical solutions (filled symbols) of $\varphi$ along the $x$-direction in the Gaussian-type density field described in Eq.~(\ref{eq:gaussian_dens}), and the analytical results of Eq.~(\ref{eq:varphi_gaussian_dens}) (solid curves of the same colour as the symbols). For this test the simulation box size is $250h^{-1}$Mpc and $x$ is rescaled to make $x/B\in[0,1]$. The simulation mesh has $256^3$ cells.} \label{fig:test_gaussian_dens}
\end{figure}

%The solution to $\varphi$ can then be obtained analytically and is
%\begin{eqnarray}\label{eq:varphi_gaussian_dens}
%\varphi &=& \bar{\varphi}\left[1-\alpha\exp\left(-\frac{(x-0.5)^2}{W^2}\right)\right],
%\end{eqnarray}
%which clearly shows that when $\alpha\rightarrow1$ $|\varphi|$ could be made very small at $x=0.5$ while at $x\rightarrow0$ or $x\rightarrow1$ it goes to $\bar{\varphi}$.

Fig.~\ref{fig:test_gaussian_dens} shows the test results for 6 out of the 7 models summarised in Table~\ref{tab:test_models} at $a=1.0$, and again the numerical solutions (symbols) match the analytical solutions (solid curves) of Eq.~(\ref{eq:varphi_gaussian_dens}) ver accurately.

%We have implemented Eq.~(\ref{eq:gaussian_dens}) into our numerical code and the numerical solutions for $\varphi$ are shown in Fig.~\ref{fig:test_gaussian_dens}. We can see that they agree with the analytical solution Eq.~(\ref{eq:varphi_gaussian_dens}) very well.

\subsection{Multilevels}

\begin{figure}
\includegraphics[scale=0.36]{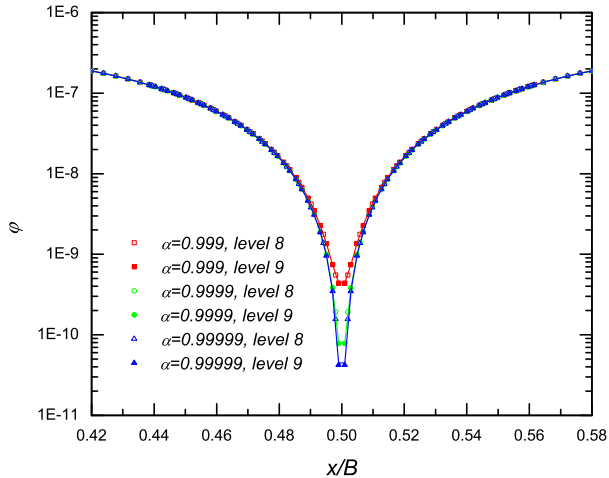}
\caption{(Colour online) Multilevel tests for model a at $a=1.0$ and three different values of $\alpha$: $0.999$ (red), $0.9999$ (green) and $0.99999$ (blue) from top to bottom. The open and filled symbols, with same shapes and colours, represent respectively the solutions to $\varphi$ along the $x$-direction on level 8 (the domain grid) and level 9 (refinement), and the corresponding analytical solutions of Eq.~(\ref{eq:varphi_gaussian_dens}) are shown using solid curves of the same colours. For this test the simulation box size is $250h^{-1}$Mpc and $x$ is rescaled to make $x/B\in[0,1]$ (code unit). The level-8 simulation mesh has $256^3$ cells.} \label{fig:test_multilevel}
\end{figure}

One of the most important features of the {\sc ecosmog} code is that it enables adaptive mesh refinements, and to make sure that this part of the code also works correctly we need to test it on the refinements. This is the task of this subsection.

%The above tests show that our code for the chameleon equation actually works accurately on the regular domain grid, but this equation is also solved on refinements where it can take different forms due the refinement boundaries. It is therefore necessary to test the chameleon equation solver on the refinements as well, as we shall do in this subsection.

The Gaussian test described above provides a good starting point for the multilevel test here, because the density contrast at $x=0.5$ could be made very large by choosing appropriate values for $\alpha$, which triggers refinements of the simulation meshes. Indeed, when $\alpha\rightarrow1$, the fast change of density field close to $x=0.5$ makes refinements essential to guarantee the high precision. For simplicity, in the test here we only refined the grid once, making this a `two-level problem', with level 8 (9) representing the coarse mesh (refinement), where `level 8' means the mesh has $2^8=256$ cells in each dimension. On both levels we used Eq.~(\ref{eq:gaussian_dens}) to set the density values in cells, and for level 9 we set the boundary conditions for $\varphi$ by interpolating the corresponding values in the coarse cells which cover the refinement boundary (more details can be found in \cite{ecosmog}).

%The Gaussian-type density configuration could provide a good check of the multilevel chameleon equation solver because the density peak could be made arbitrarily high by adjusting the parameter $\alpha$. In the vicinity of its peak, the density field changes rapidly and a higher spatial resolution is needed to compute $\varphi$ accurately. Consider the case where the regular domain grid is refined only once, in regions where the density value exceeds some certain threshold (we shall call this a 'two-level problem', and in the present numerical example the coarse and fine levels are respectively levels 8 and 9). The density values in both the coarse and refined cells are obtained by Eq.~(\ref{eq:gaussian_dens}), while the values of $\varphi$ at the fine-level boundaries are computed from interpolation of those in the nearby coarse-level cells, as discussed in \cite{ecosmog}.

Fig.~\ref{fig:test_multilevel} shows the test results for model a only and for 3 different values of $\alpha$ ($0.999$, $0.9999$ and $0.99999$ from top to bottom). In each case, we represent the numerical solutions on levels 8 and 9 by open and filled symbols of the same shape and colour, and the analytical results Eq.~(\ref{eq:varphi_gaussian_dens}) by solid curves of the same colour. Not surprisingly, the numerical and analytical solutions agree very well; so do the numerical solutions on the two different levels.

%shows the numerical values of $\varphi$ on both levels in the region covered by the refinement. We, and for each $\alpha$ the results from the coarse and fine levels are denoted respectively by empty and filled symbols. For comparison we have also plotted the analytical results Eq.~(\ref{eq:varphi_gaussian_dens}) as solid curves. As we can see, both fine-level and coarse-level results are virtually indistinguishable from the exact solution.
%
%This does not mean that the refinement is unnecessary however, because, as shown in Fig.~\ref{fig:test_multilevel}, the fine level has more data points and could probe regions closer to the extreme value of $\varphi$, which corresponds to the high-density region where high resolution is needed.

These tests make us confident about the reliability of our code, and about the simulations we describe below.

\section{Cosmological Simulations}

\label{sect:simulations}

\begin{figure*}
\includegraphics[scale=0.6]{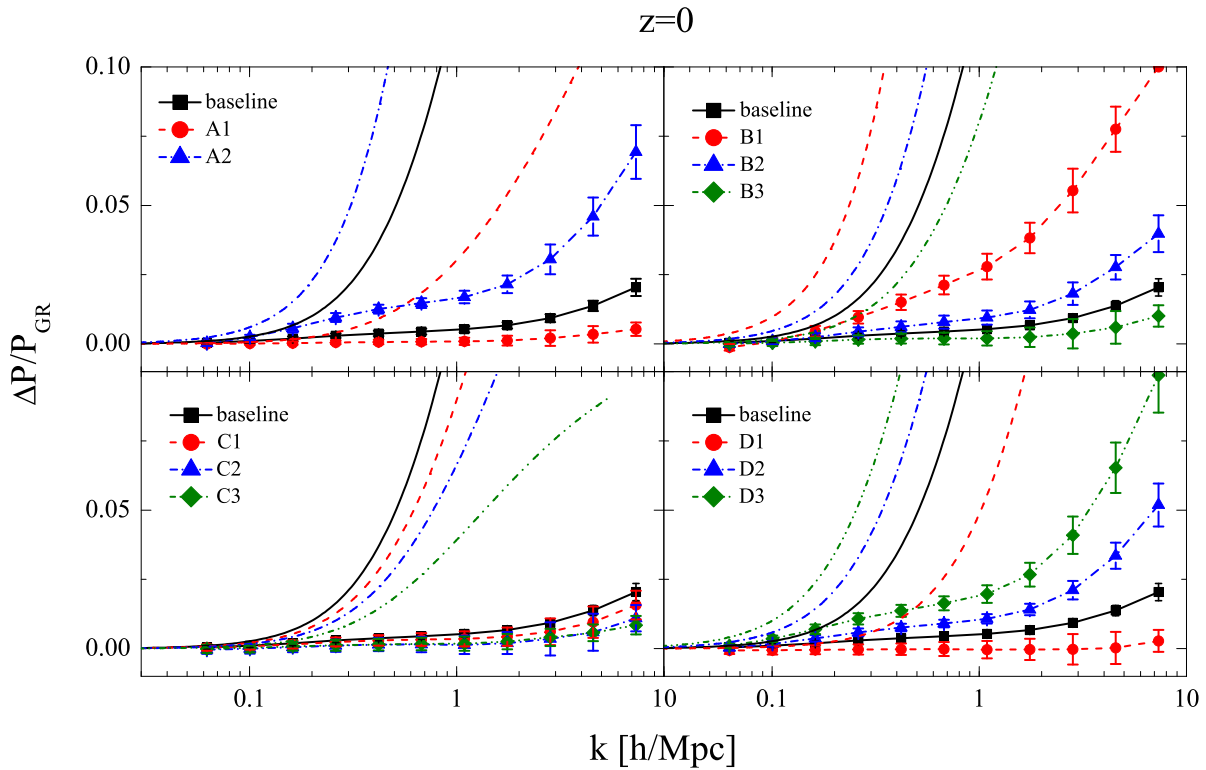}
\caption{The fractional difference in matter power spectra of various chameleon models (different models are illustrated in the legend) with respective to that of the $\Lambda$CDM model at $z=0$. The curves with error bars show the simulation result, while the curves without error bars stand for the linear theory prediction. In each panel, the curves with the same color and line style represent the same chameleon model.} \label{fig:pk_z0}
\end{figure*}

\begin{figure*}
\includegraphics[scale=0.6]{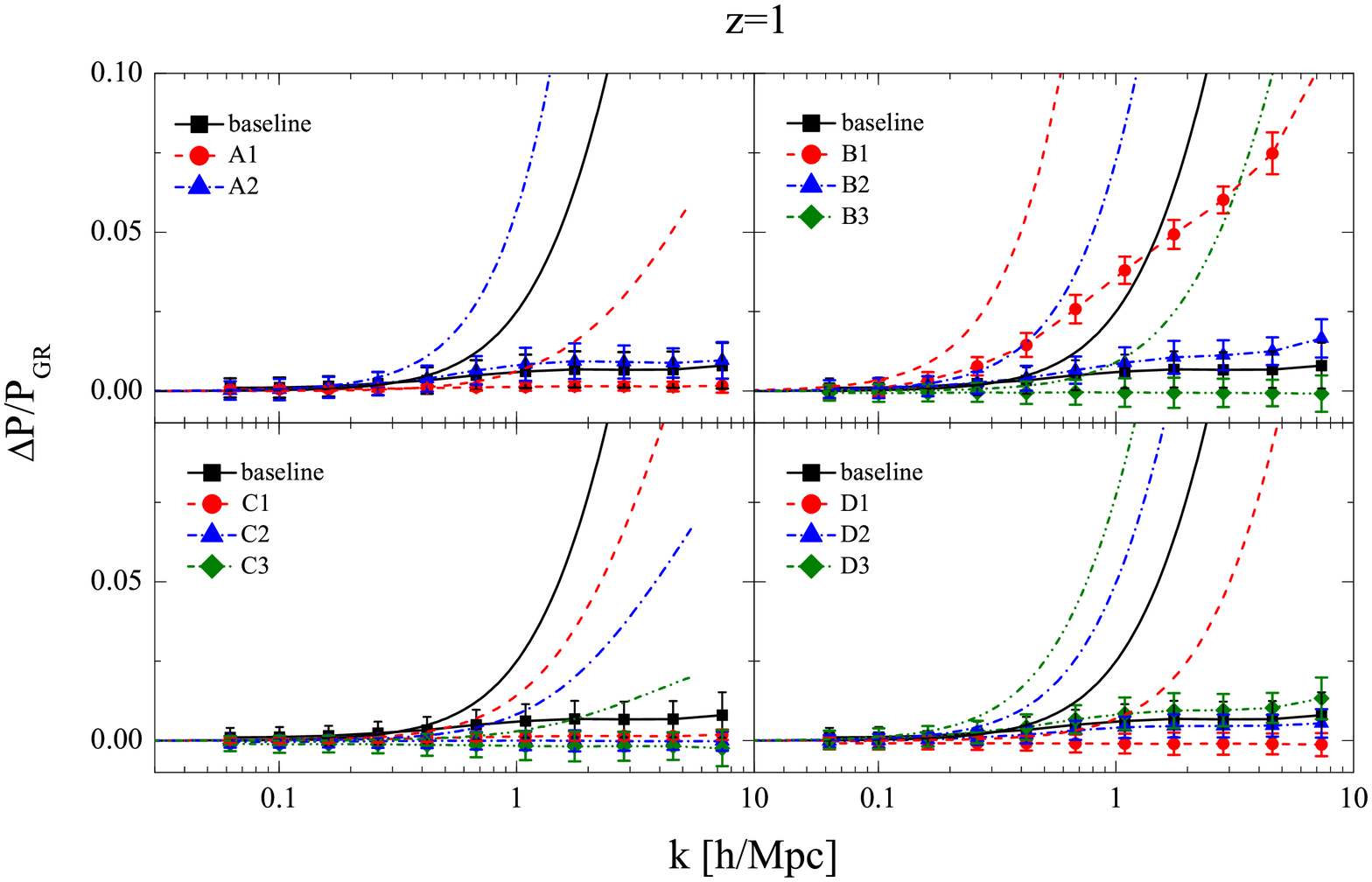}
\caption{The same as Fig \ref{fig:pk_z0}, but at $z=1$.} \label{fig:pk_z1}
\end{figure*}

\subsection{Simulation details}

This section is the core of this paper, and it shows the results of the cosmological simulations of our generalised chameleon models. We simulated a total 13 models with different values of $\beta_0$, $r$, $s$ and $\xi$, including the special case of $\Lambda$CDM which corresponds to $\beta_0=0.0$, as summarised in Table~\ref{tab:models_chameleon}. For each of the models we have 5 realisations to be averaged over to make the physical predictions more statistically meaningful, and all these 5 realisations have the same physical and simulation parameters, with the only difference being in their initial conditions, which are generated by {\sc mpgrafic} \cite{mpgrafic} at an initial redshift $z_i=49.0$ using different seeds of random numbers.

%For each model we run 5 realisations with the same physical parameters and simulation specifications, but different realisations of the initial conditions. The initial conditions are generated using {\sc mpgrafic} \citep{mpgrafic} at redshift $z_i=49.0$ with different seeds of random numbers. Since at $z_i=49.0$ the effect of the fifth force is negligible, the initial conditions should be the same for all models studied here. For the ease of comparison, we use the same random seed to generate initial conditions for the same realisation of all models, including chameleon and $\Lambda$CDM.

The cosmic expansion rate in all our simulated chameleon models is very close to that of the standard $\Lambda$CDM paradigm \citep{bdlw2012}, and this is determined by the WMAP7 \cite{wmap7} cosmological parameters. In particular,
\begin{eqnarray}
\big\{h, \Omega_m, \Omega_\Lambda, n_s, \sigma_8\big\} &=& \{0.71, 0.267, 0.733, 0.963, 0.801\}.\nonumber
\end{eqnarray}
Our simulation box is 128$h^{-1}$Mpc in each dimension, and the domain grid\footnote{In AMR codes such as {\sc ramses} and {\sc ecosmog}, the domain grid is the uniform (regular) grid which covers the whole simulation domain.} has $256^3$ cubic cells. Any cell is refined and split into 8 son cells when the number of particles inside it exceeds 9.0, and in our simulations the finest refinement level has $2^{14}$ cells on each side assuming that it covers the whole box. We use $N_p = 256^3$ dark matter particles in the simulations.

%The chameleon models are specified by the four model parameters $\beta_0$, $r$, $s$ and $\xi$. In Table~\ref{tab:models_chameleon} we list the parameters for the 13 models we have simulated.

In the rest of this subsection, we will focus on the effects of changing each model parameter on the major cosmological observables such as the matter power spectrum and halo mass function.
More specifically, we will analyse the results of our numerical simulations according to the following:
\begin{enumerate}
\item How the amplitude of the coupling strength today, $\beta_0$, affects the results: models A1 and A2;
\item How the power of the scalar field mass, $r$, affects the results: models B1, B2 and B3;
\item How the power of the coupling strength, $s$, affects the results: models C1, C2 and C3;
\item How the range of the fifth force, $\xi$, influences the results: model D1, D2 and D3.
\end{enumerate}
To see more clearly the effect of varying these four parameters, we have also simulated a baseline model $\{\beta_0,r,s,\xi\}=\{0.50,3.0,0,0.001\}$, to which all other models are compared.

\begin{table}
\begin{tabular}{@{}lcccccc}
\hline\hline
model name & $\beta_0$ & $r$ & $s$  & $\xi$ & realisations\\
\hline
$\Lambda$CDM\ \ & --\ \ & --\ \  & --\ \  & --\ \ & $5$\\
baseline\ \   & $0.50$\ \   & $3.0$\ \ & $0$\ \ & $0.001$\ \ & $5$\\
A1\ \   & $0.25$\ \   & $3.0$\ \ & $0$\ \ & $0.001$\ \ & $5$\\
A2\ \   & $0.75$\ \   & $3.0$\ \ & $0$\ \ & $0.001$\ \ & $5$\\
B1\ \   & $0.50$\ \   & $2.0$\ \ & $0$\ \ & $0.001$\ \ & $5$\\
B2\ \   & $0.50$\ \   & $2.5$\ \ & $0$\ \ & $0.001$\ \ & $5$\\
B3\ \   & $0.50$\ \   & $3.5$\ \ & $0$\ \ & $0.001$\ \ & $5$\\
C1\ \   & $0.50$\ \   & $3.0$\ \ & $-0.25$\ \ & $0.001$\ \ & $5$\\
C2\ \   & $0.50$\ \   & $3.0$\ \ & $-0.5$\ \ & $0.001$\ \ & $5$\\
C3\ \   & $0.50$\ \   & $3.0$\ \ & $-1$\ \ & $0.001$\ \ & $5$\\
D1\ \   & $0.50$\ \   & $3.0$\ \ & $0$\ \ & $0.0005$\ \ & $5$\\
D2\ \   & $0.50$\ \   & $3.0$\ \ & $0$\ \ & $0.0015$\ \ & $5$\\
D3\ \   & $0.50$\ \   & $3.0$\ \ & $0$\ \ & $0.002$\ \ & $5$\\
\hline\hline
\end{tabular}\label{tab:models_chameleon}
\caption{The parameter values for the 65 cosmological simulations we have performed for this study. Note that `--' means that the parameters are unused for the $\Lambda$CDM case.}
\end{table}

\subsection{Nonlinear matter power spectra}

The most direct way to see the effect of modified gravity on the clustering of matter is to look at the matter power spectrum $P(k)$. For this we have measured the $P(k)$ for our generalised chameleon theories and the $\Lambda$CDM paradigm using  {\sc powmes} \cite{powmes}, and calculated the relative difference $\Delta P/P_{\rm GR}$. The results are shown in Figs.~\ref{fig:pk_z0} and \ref{fig:pk_z1}.

In Figs.~\ref{fig:pk_z0} and \ref{fig:pk_z1}, we can see that both the linear perturbation results (the smooth curves) and the simulation predictions (symbols) follow the trend as we have expected (see \S~\ref{subsect:effect_param}). \tcr{The linear perturbation prediction significantly overestimates  the relative growth with respect to that in $\Lambda$CDM model in all cases, similar to what we found in the dilaton, symmetron and $f(R)$ gravity simulations \cite{bdlwz2012,lhkzjb2012}. In particular, we notice from these figures that, linear perturbation theory fails whenever it predicts a deviation from $\Lambda$CDM, and this can happen on scales as large as $k\sim0.05$Mpc$^{-1}$. This result casts strong doubts on all the efforts which have been made to constrain chameleon-type theories using linear theory predictions, and shows once again the crucial role played by nonlinear simulations.} 

\tcr{It may seem to be surprising that linear theory breaks down on large scales which can be well described by it in standard cosmology, and the reason is that the chameleon theory itself is nonlinear, and this nonlinearity is in addition to the usual nonlinearity in real matter distributions. Consequently, in the fifth-force calculation different Fourier modes of the density field strongly couple and the fifth force for large-scale modes depends on the matter perturbation on smaller scales. In linear perturbation theory, such mode coupling has been suppressed. Another way to understand the point is the following: in linear theory, the fifth force (is assumed to) depend only on the background matter density, while in nonlinear simulations it actually depends on the matter density inside overdensities which is generally higher than the background density and therefore makes it more suppressed due to the chameleon mechanism.}

%the chameleon mechanism is a nonlinear phenomenon and works most efficiently in high-density regions (with matter density perturbations higher than a few hundred); but linearising the equations (in which case the matter density perturbations are supposed to be much smaller than unity), one artificially removes or weakens this nonlinearity.

Note that the agreement between linear perturbation theory and $N$-body simulation results is up to smaller scales at $z=1$ than at $z=0$, but this is most likely because both approaches predict smaller deviations from $\Lambda$CDM at earlier times when the matter density is higher overall, rather than because linear theory works better at higher redshifts when density perturbations are small. Indeed, a direct comparison between Figs.~\ref{fig:pk_z0} and \ref{fig:pk_z1} confirms that the (nonlinear) chameleon effect is much stronger at early times.

The upper left panel of Fig.~\ref{fig:pk_z0} shows the effect of varying $\beta_0$ while all the other parameters are fixed to their baseline values (c.f.~Table~\ref{tab:models_chameleon}). As shown, $\Delta P/P_{\rm GR}$ increases when $\beta_0$ rises. Specifically, we increase and decrease $\beta_0$ around $0.5$ (which is the value of the baseline model) by 50\% in models A1 and A2 respectively, and find strong variations in the linear theory predictions of $\Delta P/P_{\rm GR}$. The simulation result of $\Delta P/P_{\rm GR}$, however, is smaller than $\sim2\%$ down to $k=0.1h$Mpc$^{-1}$ even at $z=0$. This small deviation is beyond the precision of all current cosmological probes. Recall that $\beta_0$ here is chosen to have the same value as that in the dilaton simulations of \cite{bdlwz2012}, where $\Delta P/P_{\rm GR}$ can be more than $\sim30-40\%$ -- this shows clearly that the chameleon screening is much more efficient in restoring GR in dense regions.

The upper right panel of Fig.~\ref{fig:pk_z0} shows the effects of varying $r$ while other parameters are all fixed to the baseline values. The result is again consistent with the analysis in \S~\ref{subsect:effect_param}, namely, $\Delta P/P_{\rm GR}$ grows as $r$ drops because a smaller $r$ means a less massive scalar field in the past or, thanks to the tomography mapping, in dense regions.
%Note that the chameleon works equally well at present for models B1, B2 and B3, but the situation is different in the past.
For example, the chameleon screening in model B1 is less efficient than that in B3 at $z>0$, making gravity relatively stronger in the former during most of the the evolution history, which is why the accumulated effect on matter clustering is much more significant in B1.

The lower left panel of Fig.~\ref{fig:pk_z0} illustrates the effect of varying $s$ while other parameters are fixed to the baseline values. As expected, $\Delta P/P_{\rm GR}$ drops as $s$ decreases, which is because a smaller coupling in the past or in dense regions necessarily means a weaker fifth force and therefore a decrease in the matter clustering. As we mentioned above, to avoid the unwanted anti-chameleon effect we have to choose $s\leq0$, which means that the baseline model, with $s=0$, gives the {\it largest possible} deviation from $\Lambda$CDM, which is $\lesssim1\%$ at $k=0.1h$Mpc$^{-1}$ -- this clearly implies that $s$ is practically unconstrained except that $s\leq0$.

Finally, in the lower right panel of Fig.~\ref{fig:pk_z0} we have shown the effect of varying $\xi$ with all other parameters fixed. Since $\xi$ is inversely proportional to $m_0$, an increase in $\xi$ results in a smaller scalar field mass throughout the evolution history and therefore more structures form due to the weaker suppression of the fifth force. This is exactly what we see in this panel.

Overall, Figs.~\ref{fig:pk_z0} and \ref{fig:pk_z1} indicate that observational data on the matter clustering at present and in the near future will hardly place any strong constraints on the chameleon-type modified gravity theories. One therefore has to look at other cosmological probes, such as the halo mass functions and void properties, to detect any observable signatures of these theories. We will study the former in the next subsection and leave the latter to future work.

\subsubsection{Comparison with $f(R)$ gravity model}

Note that the models we study in this work generally have a much stronger chameleon effect compared to the $f(R)$ models simulated in \cite{zlk2011,lhkzjb2012}, which are the Hu-Sawicki model \cite{hs2007} with $n=1$ and $|f_{R0}|=10^{-6}, 10^{-5}, 10^{-4}$ respectively\footnote{For more details of the models and the definitions of $f_{R0}$ and $n$, see \cite{hs2007} or \cite{zlk2011,lhkzjb2012}. Here we will quote the results rather than give a thorough review.}. From Eqs.~(12, 13, 18) of \cite{zlk2011}, it is straightforward to find
\begin{eqnarray}
\frac{m}{H_0} &=& \sqrt{\frac{\Omega_m}{2|f_{R0}|}}\frac{\Big(a^{-3}+4\frac{\Omega_\Lambda}{\Omega_m}\Big)^{3/2}}{\Big(1+4\frac{\Omega_\Lambda}{\Omega_m}\Big)}.
\end{eqnarray}
From this expression we can immediately learn two things. First, the $\xi$ parameter in the Hu-Sawicki $f(R)$ model is given by
\begin{eqnarray}
\xi\ =\ \frac{H_0}{m_0}\ =\ \sqrt{\frac{2|f_{R0}|}{\Omega_m+4\Omega_\Lambda}}.
\end{eqnarray}
Taking $\Omega_m=0.25$, $\Omega_\Lambda=1-\Omega_m=0.75$ and $f_{R0}=-10^{-6}$, we have $\xi\approx0.78\times10^{-3}$. Second, $m(a)$ is a power-law function
\begin{eqnarray}
m(a) &\propto& a^{-4.5},
\end{eqnarray}
with $r=-4.5$ for $a^{-3}\gg3$, while for $a^{-3}\sim\mathcal{O}(1)$ then $m(a)$ stays almost a constant. In addition to these, it is well known that $f(R)$ gravity is a special case of chameleon theories with $\beta_0=1/\sqrt{6}$ and $s=0$.

Judging form the values of $\xi, s$ and $\beta_0$, it may seem that the Hu-Sawicki model with $f_{R0}=-10^{-6}$ should lead to smaller deviation from $\Lambda$CDM than the baseline model. It looks even more so if one considers that $r=-4.5<-3$ for small $a$, and this seems to be inconsistent with the simulations. Note here, however, that $r=-4.5$ only happens for $z\gg1$ when the fifth force is negligible anyway, and at $z\lesssim1~$ $m$ stays around $m_0$ so that the fifth force is indeed less suppressed than in the baseline model here.

%On the other hand, the models simulated here are more directly comparable to $f(R)$ models with a leading correction to GR as $1/R^\alpha$ \cite{lb2007, bbds2008}, where the corresponding $r$ parameter is given as $r= 3(\alpha+2)/2$. The case of $r=3$ can be understood as the limit $\alpha\ll 1$, which is exactly the case studied in \cite{lb2007}. This form of $f(R)$ gravity model has motivated specific types of  chameleon potentials, which are studied in detail using $N$-body simulations in \cite{lz2009, lz2010}. It can be checked that the simulation with $\gamma=0.5$ and $\mu=10^{-6}$ in \cite{lz2010} actually gives a deviation of the matter power spectrum from $\Lambda$CDM which is comparable to that of our baseline model here.

%The mass $m_0$ is such that the scalar range is  given by
%\be
%\xi=\sqrt{\frac{(n+1)\vert f_{R_0}\vert}{4\Omega_{\Lambda 0}+ \Omega_{m0}}}
%\ee
%which correspond to $\xi\sim 10^{-3}$ when $|f_{R0}|=10^{-6}$. This explains why cases with $r\ge 3$, i.e. B1 and  D1-3, show similar results as the $f(R)$ models in the large curvature regime.

\subsection{Dark matter halo mass functions}

\begin{figure*}
\includegraphics[scale=0.6]{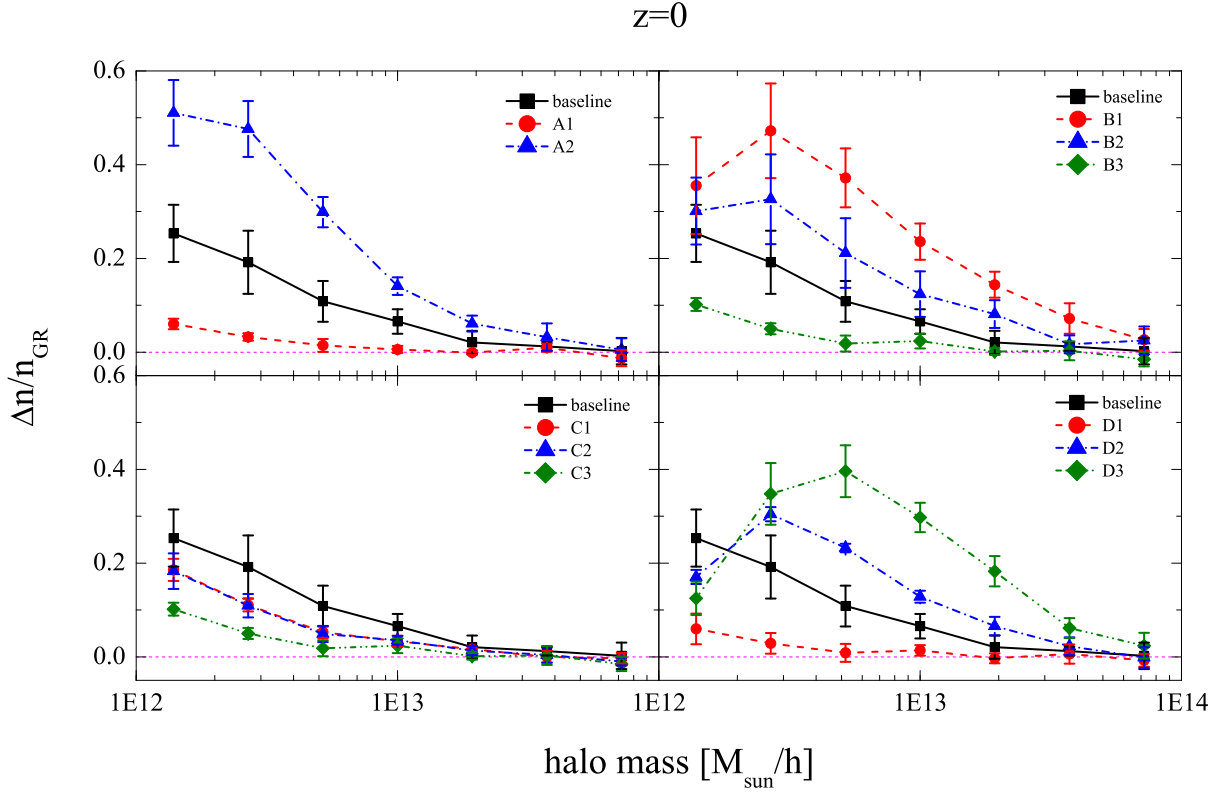}
\caption{The fractional difference in halo mass function of various chameleon models (different models are illustrated in the legend) with respective to that of the $\Lambda$CDM model at $z=0$.} \label{fig:mf_z0}
\end{figure*}

\begin{figure*}
\includegraphics[scale=0.6]{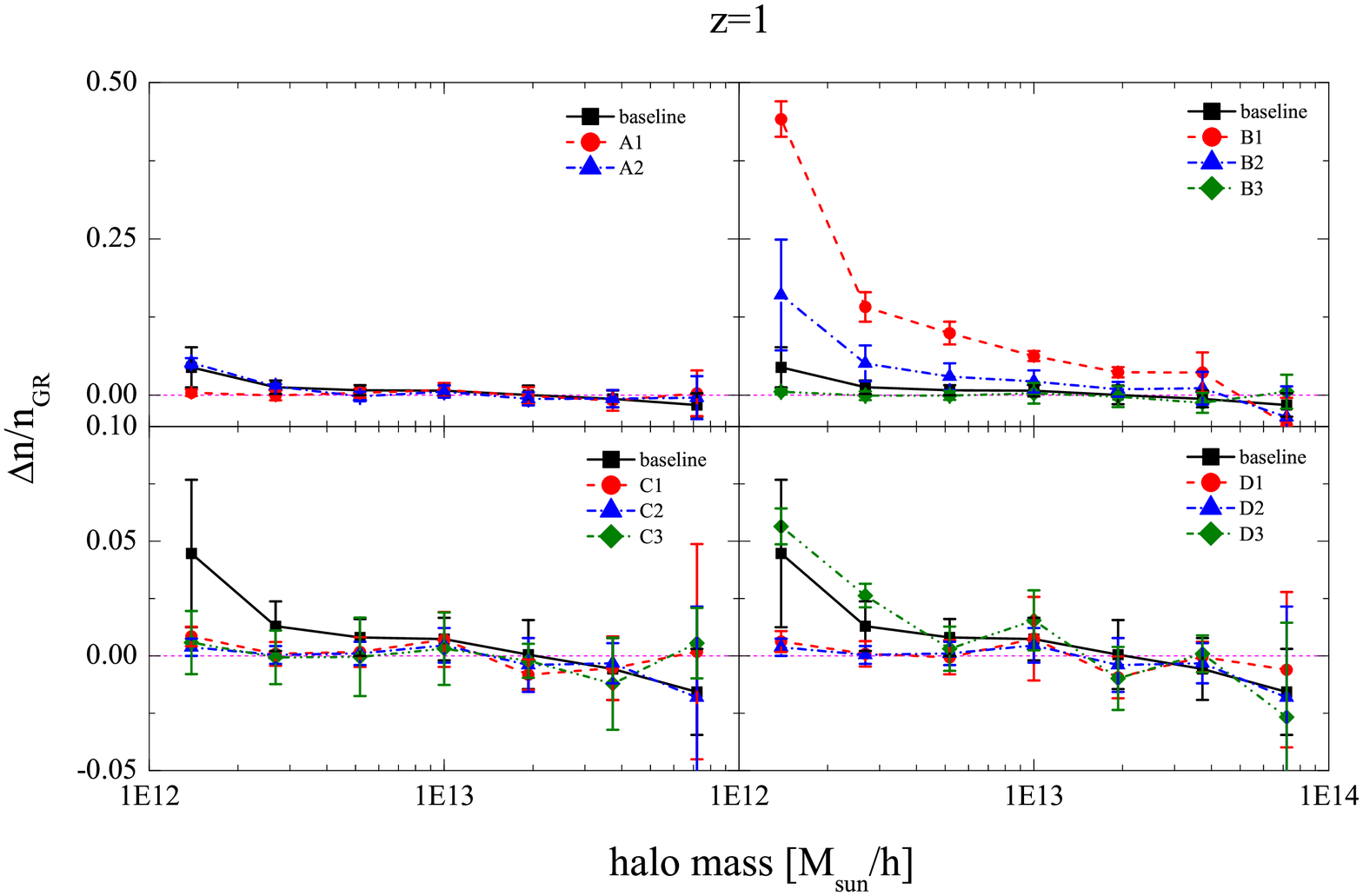}
\caption{The same as Fig \ref{fig:mf_z0}, but at $z=1$.} \label{fig:mf_z1}
\end{figure*}

We measured the dark matter halo mass functions from our simulations using the publicly available code {\sc ahf} \cite{ahf}, which is efficiently parallelised using {\sc mpi} and {\sc openmp}. We define the halo mass as the total mass contained in $R_{200}$, the radius at which the average matter density inside drops below 200 times the critical density. For each model, we have calculated the binned relative difference in mass function with respect to that of the $\Lambda$CDM model (see \cite{bdlwz2012} for details).

In Figs.~\ref{fig:mf_z0} and \ref{fig:mf_z1} we show the ratios between the chameleon and $\Lambda$CDM mass functions from our simulations at $z=0$ and $z=1$ respectively. From these figures it can be seen clearly that the fifth force leads to an overall enhancement of the formation of dark matter halos, and the effect is stronger on the low-mass end of the mass function. The maximum $\Delta n/n_{\rm GR}$ is around 50\% for the models we simulated. At $z=0$, halos with mass $M \gtrsim 5\times10^{13}h^{-1}M_{\odot}$ are generally well screened, while at $z=1$, smaller halos with mass $M \gtrsim 10^{13}h^{-1}M_{\odot}$ can also be well screened in some, if not all, cases.

The effects of varying the different chameleon parameters are generally the same as what we have expected or have seen in the plots of $\Delta P/P_{\rm GR}$, namely, $\Delta n/n_{\rm GR}$ increases as $\beta_0, s$ and $\xi$ increases or $r$ decreases. Different from the case of matter power spectra, however, the mass functions in chameleon theories show larger deviations from that of $\Lambda$CDM, particularly in the low-mass end.

A nontrivial feature in Fig.~\ref{fig:mf_z0} is the turnover on $\Delta n/n_{\rm GR}$ for models B1, B2, D1 and D2. Without loss of generality, let us take model D2 as an example and compare to the $f(R)$ model with $|f_{R0}|=10^{-6}$ (F6) simulated in \cite{zlk2011}. In both cases, the largest halos in the simulation box are well-screened, both by themselves and by their environment (because large halos tend to be produced out of very dense regions). When the halo mass decreases, the self-screening becomes weaker and the halo has a higher probability of living in average, or even underdense, regions -- the weakened screening means more matter clustering and production of more halos. Of course, there is a limited supply of matter to be incorporated into halos, and when more large halos are formed there will be fewer small halos surviving the mergers and accretions, that has caused the turn-over. This is the same as what is found for the F6 model in \cite{zlk2011} (see Fig. 11 therein) and also complies with the analytical results of \cite{le2012,ll2012}.

%$\Delta n/n_{\rm GR}$ tends to grow as mass increases if chameleon doesn't work. This is why we see an increasing trend at the low mass end. However, the self-screening of the large halos efficiently suppresses $\Delta n/n_{\rm GR}$ on the large mass end, leaving a pile-up of the medium-sized halos.

The chameleon effect in the rest of our simulated models is too strong so that even low mass halos get screened to a certain extent, making the growing trend with mass at the low mass end disappear. This can be seen by looking at the D2 model in Fig \ref{fig:mf_z1}: the turnover disappears simply because the chameleon is more efficient at higher redshifts. Also note that at $z=1$ the suppression of the fifth force is so strong that the deviation from $\Lambda$CDM almost vanishes for most models, which is the same as we have seen in the $\Delta P/P_{\rm GR}$ plots above.

\section{Summary and Conclusions}

\label{sect:summary}

To summarise, in this paper we have brought together two essential techniques for the systematic studies of the nonlinear structure formation in generic modified gravity theories of the chameleon type: a simple parameterisation scheme which covers all known chameleon theories using only four parameters and a modified version of the {\sc ecosmog} code to run high-resolution simulations efficiently. This allows us, for the first time, to get an overall picture about the behaviour of general chameleon-type theories and the part of its parameter space which is relevant for cosmology.

The powerful tomography mapping \cite{bdl2011,bdlw2012} enables us to characterise the chameleon theory and its generalisations using only a few parameters. In our case, there are two parameters describing the present value of the scalar field mass ($\xi$) and its time evolution ($r$), and another two parameters describing the current value of the coupling strength ($\beta_0$) and its time evolution ($s$). These 4 parameters cover most chameleon theories studied in the literature \cite{bdl2011}, and also the cases with varying (field-dependent) coupling to {\it nonlinear} structure formation  which  have not been thoroughly investigated so far.

Following the logic of \cite{bdlwz2012}, here we focus on the qualitative and quantitative behaviour of the generalised chameleon theory. We are interested not only in how varying the parameters changes the predictions of cosmological observables, but also in how large the changes could be such that we can decide which portion of the 4D parameter space would be of interest to cosmologists and therefore merits further (and more detailed) investigations in the future. As a by product, we want to compare the efficiencies of the different screening mechanisms that have been explored by theorists -- the chameleon, dilaton and symmetron mechanisms.

To this end, we have simulated a total of 12 models which form an extensive span in the parameter space. Starting from a default model with $\{\beta_0, r, s, \xi\} = \{0.5, 3.0, 0.0, 0.001\}$, we let each of the 4 parameters vary and take a few different values as summarised in Table~\ref{tab:models_chameleon}. In this way, we can see clearly the effect of changing every parameter.

The simulation results confirm our qualitative predictions based on simple physical arguments, namely the the fifth force (and therefore the clustering of matter) is stronger if one:
\begin{enumerate}
\item increases $\beta_0$, which results in an overall increase in the coupling strength between matter and the scalar field;
\item increases $s$, which makes the coupling strength reduce more slowly as the matter density increases;
\item increases $\xi$, which increases the range of the fifth force overall, or
\item decreases $r$, which makes the fifth force less exponentially suppressed in high-density regions.
\end{enumerate}

There are a few noticeable features which can be seen from the nonlinear matter power spectrum predicted by our simulations. The first is that, as in the cases of dilaton \cite{bdlwz2012} and $f(R)$ gravity \cite{lhkzjb2012} models where the screening is strong, linear perturbation theory fails for general chameleon theories wherever it predicts a deviation from $\Lambda$CDM. The scale at which linear theory breaks down can be as large as $k\sim0.05~h$Mpc$^{-1}$: this is typically the scale where it is assumed to be valid. This casts doubts about the reliability of the works in which linear theory predictions are used to constrain modified gravity theories such as chameleon, dilaton, symmetron and $f(R)$ gravity.

Another feature of the chameleon theory is its efficiency of screening. The model parameters here, such as $\beta_0$ and $\xi$, are chosen to be roughly the same as those in our previous dilaton and symmetron simulations \cite{bdlwz2012}, but whilst the nonlinear matter power spectra in those models can differ from those in $\Lambda$CDM by more than $30-40$\%, chameleon theories generally predict much smaller deviations ($\lesssim10\%$), indicating that the chameleon screening could restore GR much more easily. For the same reason, the effect of the fifth force also diminishes more quickly backwards in time, compared to the symmetron and dilaton cases \cite{bdlwz2012} -- indeed at redshift $z=1$ the fifth force is almost completely screened in all our simulated models except for B1, which has $r=2.0$, meaning that the scalar field mass $m$ increases more slowly with matter density. The result implies that the strength of the fifth force is very sensitive to $r$, which is, of course, as expected.

Similar features can also be seen from the dark matter halo mass functions. Here we find that, compared with the dilaton and symmetron theories \cite{bdlwz2012}, the deviations from $\Lambda$CDM are more suppressed in the high-mass end, which can be because large halos are more efficient in self-screening and also tend to be more screened by the environment because they are more likely to live in high-density environments. This is qualitatively similar to what we see in $f(R)$ gravity simulations \cite{zlk2011}. Notice that the time evolution of the halo mass function shows the same pattern as the nonlinear matter power spectra, namely that at $z=1$ the deviation from $\Lambda$CDM is very small.

The high efficiency in chameleon screening means that our choices of the parameter values might be too conservative: a deviation from the $\Lambda$CDM matter power spectrum of $\lesssim10\%$ can hardly be detected with precision in the near future, especially because the deviations are mostly on small scales where baryonic physics and other effects could already be important. Consequently, we think that future simulations of chameleon-type theories should be done for less conservative choices of parameters, namely larger values of $\beta_0, s, \xi$ and smaller values for $r$. We hope that this work can serve as a useful guidance for such future works.

\begin{acknowledgments}
ACD is supported in part by STFC. BL acknowledges supports by the Royal Astronomical Society and Durham University. HAW thanks the Research Council of Norway FRINAT grant 197251/V30 for support and Durham University for the hospitality where part of this work was carried out. GBZ is supported by a Dennis Sciama Fellowship at the University of Portsmouth. PB is partially supported by ANR BLANC 2010 041301. The simulations and the post-process of the simulation data were performed on the {\sc sciama} machine at the University of Portsmouth and on the {\sc cosma} supercomputer at Durham University.
\end{acknowledgments}

\appendix

%\section{Binning $P(k)$ and the mass function}\label{sec:make_bin}
%
%To show the result of the simulation, it is useful to quote the relative difference of the power spectrum $P(k)$ and halo mass function for the modified gravity model (chameleon model in this work) with respect to the $\Lambda$CDM model. Let  ${\mathcal{R}}\equiv x_{\rm MG}/x_{\rm\Lambda}$, where $x$ can be $P(k)$ or $n(M)$. Then the standard deviation $\sigma_\mathcal{R}$ of $\mathcal{R}$ for each $k$ bin or mass bin is computed using the rule of error propagation,
%\begin{eqnarray}
%\left(\frac{\sigma_{\BLED{\mathcal{R}}}}{\BLED{\mathcal{R}}}\right)^2 &=& \left(\frac{\sigma_{\rm MG}}{x_{\rm MG}}\right)^2 + \left(\frac{\sigma_{\rm\Lambda}}{x_{\rm\Lambda}}\right)^2-2\BLED{\hat{\rho}}\frac{\sigma_{\rm MG}}{x_{\rm MG}}\frac{\sigma_{\rm \Lambda}}{x_{\rm\Lambda}},\
%\end{eqnarray}
%The subscripts $_{\rm MG}$ and $_{\rm\Lambda}$ denote the chameleon model and $\Lambda$CDM respectively, and $\BLED{\hat{\rho}}$ is the correlation coefficient between the mass functions of the two, i.e.,
%
%\begin{equation}
%\BLED{\hat{\rho}}=\frac{\sum\limits_{i}\left(x_{\rm MG}^i-\bar{x}_{\rm MG}\right)\left(x_{\Lambda}^i-\bar{x}_{\Lambda}\right)}{\left[\sum\limits_{i}\left(x_{\rm MG}^i-\bar{x}_{\rm MG}\right)^2\sum\limits_{i}\left(x_{\Lambda}^i-\bar{x}_{\Lambda}\right)^2\right]^{1/2}}
%\end{equation} where the sum is over different realisations and the quantity with an overbar denotes the average over all realisations.

\end{document}